\documentclass[twocolumn]{aastex63} %linenumbers]{aastex63}
\usepackage{graphicx}
\usepackage[caption=false]{subfig}
\usepackage{amsmath,amsfonts,amsthm,bm}

\usepackage{hyperref}
\usepackage{xcolor}
\usepackage{verbatim}
\definecolor{medium-blue}{rgb}{0,0,1}
\hypersetup{colorlinks, urlcolor={medium-blue}}

\definecolor{my_color}{HTML}{3a18b1}

\definecolor{new_color}{HTML}{000000}%{CF0000}% this is a maroon
\definecolor{new_black}{HTML}{000000}% this is a maroon
\usepackage{xfrac}

\newcommand\bedit[1]{\textcolor{black}{#1}}
\newcolumntype{b}{>{\color{new_color}}c}

\graphicspath{{pics/}}

\newcommand{\kms}{\ensuremath{\rm km\,s^{-1}}}
\newcommand{\cms}{\ensuremath{\rm cm\,s^{-1}}}

\newcommand{\ms}{\ensuremath{\rm m\,s^{-1}}}

\newcommand{\HarvardPhysics}{Department of Physics, Harvard University, 17 Oxford Street, Cambridge MA 02138, USA}
\newcommand{\CfA}{Center for Astrophysics $|$ Harvard \& Smithsonian, 60 Garden Street, Cambridge, MA 02138, USA}
\newcommand{\NASASaganFellow}{NASA Sagan Fellow}

\newcommand{\GenevaObservatory}{Observatoire de Gen\`eve, Universit\'e de Gen\`eve, 51 chemin des Maillettes, 1290 Versoix, Switzerland}
\newcommand{\StAndrewsPhysicsAstronomy}{Centre for Exoplanet Science, SUPA, School of Physics and Astronomy, University of St Andrews, St Andrews KY16 9SS, UK}
\newcommand{\BelfastMathPhysics}{Astrophysics Research Centre, School of Mathematics and Physics, Queen's University Belfast, BT7 1NN, Belfast, UK}
\newcommand{\PadovaPhysicsAstronomy}{Dipartimento di Fisica e Astronomia ``Galileo Galilei'', Universit{\`a} di Padova, Vicolo dell`Osservatorio 3, I-35122 Padova, Italy}
\newcommand{\CavendishLab}{Astrophysics Group, Cavendish Laboratory, J.J. Thomson Avenue, Cambridge CB3 0HE, UK}

\newcommand{\EdinburghAstronomy}{SUPA, Institute for Astronomy, Royal Observatory, University of Edinburgh, Blackford Hill, Edinburgh EH93HJ, UK}
\newcommand{\EdinburghExoplanetCenter}{Centre for Exoplanet Science, University of Edinburgh, Edinburgh, UK}
\newcommand{\INAFTorino}{INAF-Osservatorio Astrofisico di Torino, via Osservatorio 20, 10025 Pino Torinese, Italy}
\newcommand{\INAFPalermo}{INAF-Osservatorio Astronomico di Palermo, Piazza del Parlamento 1, 90134 Palermo, Italy}
\newcommand{\INAFPadova}{INAF-Osservatorio Astronomico di Padova, Vicolo dell`Osservatorio 5, 35122 Padova, Italy}
\newcommand{\INAFBrenaBaja}{Fundacion Galileo Galilei-INAF, Rambla J. A. F. Perez, 7,
E-38712, S.C. Tenerife, Spain}
\newcommand{\INAFCagliari}{INAF-Osservatorio Astronomico di Cagliari, Via della Scienza 5-09047 Selargius CA, Italy}

\newcommand{\KavliInstitute}{Kavli Institute for Cosmology, University of Cambridge, Madingley Road, Cambridge CB3 0HA, UK}

\shorttitle{Removing RV Stellar Activity Signals Using Neural Nets}
\shortauthors{de Beurs et al.}
\begin{document}

% REU: The title is the single most important element of the paper, because it is the part
% most widely used for literature searches.  
%
% Your title should be about  WHAT you did, and not tightly focused on HOW you did it.  
% The methods you used are of secondary importance to what you accomplished.  
% In other words, put your science first.  Try not to use the overused word "study" in the title.

\title{Identifying Exoplanets with Deep Learning. IV. Removing Stellar Activity Signals from Radial Velocity Measurements Using Neural Networks}

\correspondingauthor{Zoe L.\ de Beurs}
\email{zdebeurs@utexas.edu}

% REU: This shows how to format the names.  Use your full name, spelled out the way
% you want it to appear in the literature.  The 16-digit number inside the [] should
% be replaced with your ORCID.  If you don't have one, you should sign up for one,
% and put that number here and in all your future papers.  
% For your affiliation, use your home institution and address, not the CfA.
% You and your advisor should work out together whose names appear here,
% the basic criterion is that people who have made essential contributions 
% to the manuscript should be included.

\author[0000-0002-7564-6047]{Zoe\ L. de Beurs}
\affiliation{Department of Earth, Atmospheric and Planetary Sciences, Massachusetts Institute of Technology, Cambridge, MA 02139, USA}
\affiliation{Department of Astronomy, University of Texas at Austin, 2515 Speedway, Austin, Texas 78712, USA}
\affiliation{Department of Astronomy, University of Wisconsin-Madison, Madison, WI, 53706, USA}
\affiliation{NSF Graduate Research Fellow}

\author[0000-0001-7246-5438]{Andrew Vanderburg}
\affiliation{Department of Physics and Kavli Institute for Astrophysics and Space Research, Massachusetts Institute of Technology, Cambridge, MA 02139, USA}
\affiliation{Department of Astronomy, University of Wisconsin-Madison, Madison, WI, 53706, USA}
\affiliation{Department of Astronomy, University of Texas at Austin, 2515 Speedway, Austin, Texas 78712, USA}
\affiliation{\NASASaganFellow}

\author{Christopher J. Shallue}
\affiliation{\CfA}
%\nocollaboration
 
\author[0000-0002-9332-2011]{Xavier Dumusque}
\affiliation{\GenevaObservatory}

\author[0000-0002-8863-7828]{Andrew Collier Cameron}
\affiliation{\StAndrewsPhysicsAstronomy}

\author{\bedit{Christopher Leet}}
\affiliation{\bedit{Facebook, 181 Fremont St, San Francisco, CA 94105, USA}}

\author[0000-0003-1605-5666]{Lars A. Buchhave}
\affiliation{DTU Space, National Space Institute, Technical University of Denmark, Elektrovej 328, DK-2800 Kgs. Lyngby, Denmark}

%\author[0000-0001-7032-8480]{S. H. Saar}
%\affiliation{\CfA}

%\author[0000-0002-6492-2085]{L. Malavolta}
%\affiliation{\PadovaPhysicsAstronomy}

%\author{S. Thompson}
%\affiliation{\CavendishLab}

%\author[0000-0002-8863-7828]{A. Collier Cameron}
%\affiliation{\StAndrewsPhysicsAstronomy}

%\author[0000-0002-9332-2011]{X. Dumusque}
%\affiliation{\GenevaObservatory}

%% TIER TWO

%\author[0000-0001-8934-7315]{H. M. Cegla}
%\altaffiliation{\CHEOPSFellow}
%\altaffiliation{\UKRIFellow}
%\affiliation{\GenevaObservatory}
%\affiliation{\WarwickPhysics}

%\author{J. Costes}
%\affiliation{\BelfastMathPhysics}

%\author[0000-0002-2218-5689]{J. Maldonado}
%\affiliation{\INAFPalermo}

%\author{N. Buchschacher}
%\affiliation{\GenevaObservatory}

%\author[0000-0001-5701-2529]{M. Cecconi}
%\affiliation{\INAFBrenaBaja}

%\author[0000-0002-9003-484X]{\textcolor{red}{David Charbonneau}}
%\affiliation{\CfA}

\author[0000-0003-1784-1431]{Rosario Cosentino}
\affiliation{\INAFBrenaBaja}

\author[0000-0003-4702-5152]{Adriano Ghedina}
\affiliation{\INAFBrenaBaja}

\author[0000-0001-9140-3574]{R. D. Haywood}
\affiliation{Astrophysics Group, University of Exeter, Exeter EX4 2QL, UK}
\affiliation{STFC Ernest Rutherford Fellow}

%\author{A. G. Glenday}
%\affiliation{\CfA}

%\author{M. Gonzalez}
%\affiliation{\INAFBrenaBaja}

%\author{C-H. Li}
%\affiliation{\CfA}

%\author[0000-0002-5432-9659]{M. Lodi}
%\affiliation{\INAFBrenaBaja}
\author[0000-0003-2107-3308]{Nicholas Langellier}
%\altaffiliation[Corresponding Author: ]{nlangellier@gmail.com}
\affiliation{\HarvardPhysics}
\affiliation{\CfA}

\author[0000-0001-9911-7388]{David W. Latham}
\affiliation{\CfA}

\author[0000-0003-3204-8183]{Mercedes L\'opez-Morales}
\affiliation{\CfA}

%\author{\textcolor{red}{Christophe Lovis}}
%\affiliation{\GenevaObservatory}

\author{Michel Mayor}
\affiliation{\GenevaObservatory}

\author[0000-0002-9900-4751]{Giusi Micela}
\affiliation{\INAFPalermo}

\author[0000-0001-5446-7712]{Timothy W. Milbourne}
\affiliation{\HarvardPhysics}
\affiliation{\CfA}

\author[0000-0001-7254-4363]{Annelies Mortier}
\affiliation{\CavendishLab}
\affiliation{\KavliInstitute}

\author[0000-0002-1742-7735]{Emilio Molinari}
%\affiliation{\INAFBrenaBaja}
\affiliation{\INAFCagliari}

\author{Francesco Pepe}
\affiliation{\GenevaObservatory}

\author[0000-0001-5132-1339]{David F. Phillips}
\affiliation{\CfA}

\author[0000-0002-4445-1845]{Matteo Pinamonti}
\affiliation{\INAFTorino}

\author[0000-0002-9937-6387]{Giampaolo Piotto}
\affiliation{\PadovaPhysicsAstronomy}
\affiliation{\INAFPadova}

%\author[0000-0003-1200-0473]{\textcolor{red}{Ennio Poretti}}
%\affiliation{\INAFBrenaBaja}
%\affiliation{\INAFBrera}

\author[0000-0002-6379-9185]{Ken Rice}
\affiliation{\EdinburghAstronomy}
\affiliation{\EdinburghExoplanetCenter}

\author[0000-0001-7014-1771]{Dimitar Sasselov}
\affiliation{\CfA}

%\author{\textcolor{red}{Damien S{\'e}gransan}}
%\affiliation{\GenevaObservatory}

\author[0000-0002-7504-365X]{Alessandro Sozzetti}
\affiliation{\INAFTorino}

%\author{A. Szentgyorgyi}
%\affiliation{\CfA}

\author{St\'ephane Udry}
\affiliation{\GenevaObservatory}

\author{Christopher A. Watson}
\affiliation{\BelfastMathPhysics}

%\author{R. L. Walsworth}
%\affiliation{\CfA}
%\affiliation{\MarylandPhysics}
%\affiliation{\MarylandECE}
%\affiliation{\MarylandQTC}

%\author{\textbf{+HARPS-Solar Telescope Authors}}

% Mark off the abstract in the ``abstract'' environment. 
\begin{abstract}

Exoplanet detection with precise radial velocity (RV) observations is currently limited by spurious RV signals introduced by stellar activity. We show that machine learning techniques such as linear regression and neural networks can effectively remove the activity signals (due to starspots/faculae) from RV observations. Previous efforts focused on carefully filtering out activity signals in time using modeling techniques like Gaussian Process regression (e.g. \citealt{2014MNRAS.443.2517H}). Instead, we systematically remove activity signals using only changes to the average shape of spectral lines, and no information about when the observations were collected. We trained our machine learning models on both simulated data (generated with the SOAP 2.0 software; \citealt{2014ApJ...796..132D}) and observations of the Sun from the HARPS-N Solar Telescope \citep{dumusque2015, 2016SPIE.9912E..6ZP, 2019MNRAS.487.1082C}. We find that these techniques can predict and remove stellar activity from both simulated data (improving RV scatter from 82 \cms\ to 3 \cms) and from more than 600 real observations taken nearly daily over three years with the HARPS-N Solar Telescope (improving the RV scatter from \bedit{1.753} \ms\ to \bedit{1.039} \ms, a factor of $\sim$  \bedit{1.7} improvement). In the future, these or similar techniques could remove activity signals from observations of stars outside our solar system and eventually help detect habitable-zone Earth-mass exoplanets around Sun-like stars.
\end{abstract}


\section{Introduction}
\label{sec:intro}
The Radial Velocity (RV) method has seen tremendous improvements since the first detections of exoplanets around Sun-like stars between 1988 and 1995 \citep{1995Natur.378..355M, 1989Natur.339...38L, 1988ApJ...331..902C}. Currently, the primary challenge in measuring Extremely Precise Radial Velocities (EPRVs) is overcoming noise from stellar variability \citep{national2018exoplanet, 2020arXiv200513386H}. The surfaces of Sun-like stars are affected by numerous phenomena from convective granulation to magnetic activity in the form of spots, plages and faculae. Due to the time-evolving and sometimes periodic nature of these features, they have been mistaken for planets on several occasions \citep[e.g.][]{2008Natur.451...38S, 2008A&A...489L...9H, 2001A&A...379..279Q} and can severely complicate the interpretation of RV measurements. Currently, these forms of stellar variability commonly limit RV measurement precision to $\gtrsim$ 1 \ms\ \citep{2018A&A...620A..47D, 2016MNRAS.457.3637H}. To detect the 10 \cms\ signals induced by Earth-mass exoplanets in the habitable zones of Sun-like stars, our limiting RV precision must improve by an order of magnitude.

Characterizing and removing these stellar activity\footnote{\bedit{We note that throughout the paper, we will at times refer to stellar variability and stellar activity interchangeably because the main variability contribution in our dataset is magnetic activity.}} signals is especially crucial and timely as current and future high-resolution spectrographs (including HARPS; \citealt{2003Msngr.114...20M, 2012Natur.485..611W}, HARPS-N; \citealt{2012SPIE.8446E..1VC}, ESPRESSO; Pepe et al 2020 (\textit{accepted}), G-CLEF; \citealt{2014SPIE.9147E..26S} EXPRES; \citealt{2016SPIE.9908E..6TJ}) already have \citep{2016Natur.536..437A, 2020A&A...639A..77S} or are expected to reach the long-term instrumental RV precision necessary to detect Earth-mass habitable-zone exoplanets.

The signals that limit RV precision on stars like the Sun are caused by four main physical processes:
\begin{enumerate}
\item \textit{Solar-type oscillations -} produced by pressure waves propagating at the surface, pressure-mode (p-mode) oscillations result in a contraction and expansion of the external envelope of the star on timescales of a few minutes \citep{1995A&A...293...87K, 1962ApJ...135..474L, 1970ApJ...162..993U, Butler_2003, 2008ApJ...687.1180A}. These oscillations can produce RV signals ranging from 10 \cms\ to 1 \ms\ for solar-like stars \citep{2008ApJ...687.1180A}. The period and amplitude vary depending on the stellar type and evolutionary stage. For our Sun, this RV variation is at the 0.5 \ms\ level at a $\sim$ 5 minute period \citep{2018A&A...612A..44S, 2019Geosc...9..114C}.

\item \textit{Granulation phenomena -} originating from convection in the outer layers of solar-type stars, granulation and supergranulation can induce RV signals at the \ms\ level \citep{2008A&A...490.1143L, 2011A&A...525A.140D} on timescales of a few minutes up to 48 hours. These granulation phenomena are found throughout the photosphere except in active regions where convection is suppressed by magnetic fields \citep{1990A&A...231..221B, 1982Natur.297..208L, 1981A&A....96..345D}.

\item \textit{Short-term stellar activity -} induced by stellar rotation paired with dark spots and bright faculae on the surface of the sun, short-term stellar activity is caused by two different physical effects. In the first effect, the presence of strong magnetic fields in active regions suppresses the convection and thereby the convective blueshift effect \citep{1990A&A...231..221B, 1985A&A...143..116C, 1982Natur.297..208L, 1982ARA&A..20...61D}. Relative to the quiet photosphere, the active regions then seem redshifted \citep{1985A&A...143..116C}. As the active regions come in and out of view during the rotation, they produce RV signals of $\sim 0.4-1.4$ \ms\ for the Sun \citep{meunier2010}. In the second effect, the temperature difference between these active regions and the quiet photosphere result in flux differences. For example, dark sunspots are $\sim$ 700 K cooler and thus have much lower flux than the rest of the star \citep{meunier2010}. In this way, spots break the balance between the blueshifted approaching limb and the redshifted receding limb as they pass across the stellar disk and induce RV variations that can reach 0.4 \ms\ on the Sun at high activity \citep{1997ApJ...485..319S, meunier2010}. In other cases, like young stars and M-dwarfs, dark spots can dominate the activity signals. Both the suppression of convective blueshift effect and the flux effect produce RV variations on the timescale of the rotation period.

\item \textit{Long-term stellar activity -} generated by solar-like magnetic activity cycles, long-term stellar activity variations influence RV measurements on the timescale of several years. In solar-type magnetic cycles, the filling factor of active regions increases during high-activity phases. Since the increase in magnetic field in active regions suppresses the convection (and thereby convective blueshift), these areas will be relatively redshifted (positive velocity) as the activity level rises. Thus, the activity level and RVs are positively correlated \citep{meunier2010, 2003A&A...401.1185L}. \citet{2011A&A...535A..55D} found that stars other than the Sun can also have these solar-like magnetic cycles and their corresponding long-term RV variations. For our Sun, we observe an 11-year magnetic cycle during which the sunspot number varies from zero to $\sim$ 150-200 on the visible hemisphere \citep{2015LRSP...12....4H} and large, bright magnetic regions can dominate solar RVs \citep{2019ApJ...874..107M}.
\end{enumerate}
On the Sun, all four of these phenomena contribute activity signals with comparable amplitudes. In RV analysis, these signals are often aggregated into a single measurement of the stellar activity. When approximating each source of RV variations as Gaussian noise, the total scatter in an RV observation, $\sigma_{tot}$, can be summarized as:
\begin{equation}
 \sigma_{tot} \approx \sqrt{\sigma_{phot}^2+\sigma_{ins}^2+\sigma_{magn}^2+\sigma_{gran}^2+\sigma_{p-mode}^2 }
\end{equation}

\noindent where  $\sigma_{phot}$ is photon noise, $\sigma_{ins}$ is instrumental noise, $\sigma_{magn}$ originates from both short and long-term activity, $\sigma_{gran}$ is scatter from granulation phenomena, and $\sigma_{p-mode}$ is scatter from p-modes.

To mitigate some of these forms of stellar variability, observing strategies have been developed to average out noise from granulation phenomena and p-mode oscillations. \citet{2011A&A...525A.140D, 2011A&A...525A.140D} showed that RV signals caused by these two stellar noise sources can be averaged out with longer integration times and a higher frequency of observations throughout the night (or day in the case of observing the Sun). Later, \citet{medina18} extended this strategy to evolved stars. For p-mode oscillations specifically, \citet{2019AJ....157..163C} demonstrated that fine-tuning exposure times to stellar parameters (e.g. 5.4 minutes for the Sun) can also efficiently average out p-modes down to $\sim$ 10 \cms.

Other methods of distinguishing planetary systems from stellar variability include tracing activity indicators such as $\textrm{log R}_{\textrm{HK}}^{'}$ \citep{1984ApJ...279..763N}, the Bisector Inverse Slope Span \citep{2001A&A...379..279Q}, $H\alpha$ \citep{2007A&A...474..293B, 2014Sci...345..440R}, or using statistical methods (like Gaussian Processes (GPs); \citealt{2014MNRAS.443.2517H, 2015MNRAS.452.2269R, 2017arXiv171101318J, 2018A&A...614A.133D}, Moving Average; \citealt{2013A&A...549A..48T}, tomography modelling;  \citealt{2014MNRAS.444.3220D}), or measuring the RV from stellar lines that are the least affected by stellar activity \citep{2018A&A...620A..47D, 2020A&A...633A..76C} to reduce the impact of stellar activity in RV datasets. Other methods of capturing stellar activity variations include using photometry  (the FF' method; \citealt{2012MNRAS.419.3147A}, a GP framework that extends the FF' method; \citealt{Rajpaul_2015}; \bedit{using simultaneous spectroscopy and photometry to disentangle the contribution of spots, plagues, and network regions to the RV signal \citep{Milbourne2021}}, and combining RV metrics with solar photometry to predict rotation periods; \citealt{Kosiarek_2020}). \bedit{More recently, several studies have investigated how individual spectral lines are affected differently by stellar activity \citep{Dumusque2018, Cretignier2021, Wise2022}.} 

%\bedit{Recently, several studies have investigated how individual spectral lines are affected differently by stellar activity. \citet{Dumusque2018} found that by choosing spectral lines that are less affected by stellar activity to compute the RVs, the stellar activity contribution can by decreased by a factor of two. \citet{Cretignier2021} further developed this work and found that the shallow lines which are formed deep inside the star's atmosphere are most impacted by stellar activity. Next, \citet{Wise2022} applied this line-by-line approach to HARPS observations of $\alpha$ Cen B to develop an astrophysical model to explain the line variability. }%These line-by-line methods may be able to be combined with our machine learning methods to further push RV precision.

Although methods such as the FF' method and the GP frameworks have been successfully applied to numerous datasets and detected low-amplitude planetary signals, they often require and rely on high cadence and carefully timed observations. It is often difficult to obtain such timely observations given the myriad scheduling constraints involved in running astronomical observatories. Ideally, we would employ a method that successfully addresses short-term stellar activity, but does not require high sampling or timing information that is necessary for GPs and Moving Averages. In this paper, we illustrate that machine learning (ML) algorithms have the potential to resolve both these challenges by identifying changes to the average shape of spectral lines. While this technique does require a substantial set of observations for training, densely sampled observations are not necessary.

Neural networks, a form of machine learning, have solved many complex problems in many other fields ranging from natural language processing \citep{collobert2008unified} to medicine \citep{ramesh2004artificial}. Neural networks are also gaining ground in solving astrophysical problems \citep[e.g.][]{bloom12, dominguez}, including in the field of exoplanets. Specifically, neural networks have successfully identified exoplanet transits in simulated data \citep{2018MNRAS.474..478P, 2018AJ....155..147Z} as well as classified planet candidates and false positives detected by Kepler \citep{2018AJ....155...94S, 2018ApJ...869L...7A}, K2 \citep{2019AJ....157..169D}, TESS \citep{2019AJ....158...25Y,2020A&A...633A..53O}, NGTS \citep{2019MNRAS.488.5232C}, and WASP \citep{2019MNRAS.483.5534S}.

\begin{deluxetable*}{ccc}[ht!]
\centering
%\tablenum{2}
\tablecaption{Monte Carlo Parameters for Simulated Data\label{tab:soapparameters}}
\tablewidth{1pt}
\tablehead{
\colhead{Fixed   Parameters} & \colhead{ } &
\colhead{Notes}
}
\startdata
Grid & 300 & Grid Resolution Power (NxN) \\
nrho & 20 & Resolution for spot's circumference \\
Instrument resolving power & 115000 & Resolving Power of the spectrograph (115000 for HARPS-N) \\
Radius sun & 696000 & Radius of the Sun {[}km{]} {[}1{]}\\
Radius & 1 & Simulated stellar Radius{[}Rsun{]} \\
$P_{\rm{rot}}$ & 25.05 & Rotation period {[}day{]}   25.05 for the Sun {[}1{]}\\
I & 90 & Stellar inclination angle   {[}degree{]}, 0 degree: pole on (North), 90 degrees: equator on \\
Psi & 0 & initial phase \\
$T_{\rm{star}}$ & 5778 & Effective temperature of star, 5778 for the Sun {[}1{]}\\
$T_{\rm{diff\ spot}}$ & 663 & Temperature difference between the star effective temperature and the spot {[}2{]}\\\\ %, 663 for the Sun {[}2{]} \\
Limb1 & 0.29 & linear limb darkening   coefficient (can be obtained from {[}3{]}). 0.29 for the Sun ({[}4{]},{[}3{]}) \\
Limb2 & 0.34 & quadratic limb darkening   coefficient (can be obtained from {[}3{]}). 0.34 for the Sun ({[}4{]},{[}3{]}) \\
\hline
\hline
Random Parameters &  & Notes  \\
\hline
Number of Active Regions & 0-4 & Follows a Poisson distribution with most probable value of 1 \\
Active Region Type: & Spot or faculae & Equal probability of being assigned as a spot or a plage. \\
Active Region Longitude: & 0 to 360 & Random uniform distribution between 0 to 360 degrees \\
Active Region Latitude: & -90 to 90 & Random uniform distribution between -90 to 90 degrees \\
Active Region Size: & 0.0067 to 0.090 & [In units of the stellar radius] Log uniform distribution\\
\enddata
\tablenotetext{[1]}{~~  From \protect\citet[NASA Planetary Fact Sheets][]{NASA_fact_sheet}}
\tablenotetext{[2]}{~~  From \protect\citet{meunier2010}}
\tablenotetext{[3]}{~~ From \protect\cite{Claret2011}}
\tablenotetext{[4]}{~~ From \protect\cite{Oshagh2013}}
\end{deluxetable*}

Our strategy is to use ML to identify and interpret the subtle changes to stellar spectra that are caused by stellar activity. Previously, \citet{2017ApJ...846...59D} used principal component analysis (PCA) to show that photospheric activity signals are clearly distinct from Keplerian RV shifts in simulated data.  Beyond distinguishing the two phenomena, we want to be able to predict the RV signals induced by stellar activity such that we can remove these signals and reveal smaller Keplerian signals that were previously hidden. \bedit{Some preliminary methods are now emerging to separate stellar activity signals using these spectral changes \citep{CollierCameron2021, Cretignier2022, fiesta}.} In this work, we attack the problem with machine learning and train multiple models to predict and remove stellar activity RV signals from observations of the HARPS-N Solar Telescope \citep{dumusque2015, 2016SPIE.9912E..6ZP, 2019MNRAS.487.1082C}. 

Our paper is organized as follows. In Section \ref{observation_descrip}, we describe the simulated data and real observations that serve as training sets, and in Section \ref{inputrepresentation}, we describe how we process and prepare the data to be input to our ML models.  In Section \ref{train_val_test}, we describe how the observations were divided into training, (cross-)validation, and test sets. In Section \ref{architectures} and \ref{training_proc}, we describe the ML architectures we used, including several different neural networks and our training procedure. In Section \ref{results}, we present our results. In Sections \ref{discussion} and \ref{conclusion}, we discuss the implications of these results and conclude.

%To do this, we look for shape changes of the star's average spectral line profile. We measure the average line profile by calculating the cross correlation function (CCF) between the spectrum and a template composed of delta functions at the positions of the spectral lines. Photospheric activity then results in shape changes of the CCF which we can use to predict their contribution to the RV signal.

%Deep neural networks are considered state-of-the-art for shape recognition. Therefore, we use a deep learning method that learns to predict stellar activity signals based on the subtle differences in shape of Cross-Correlation Functions (CCFs) and succesfully remove this noise from stellar activity.

\begin{figure*}
    \epsscale{1.2}
    \plotone{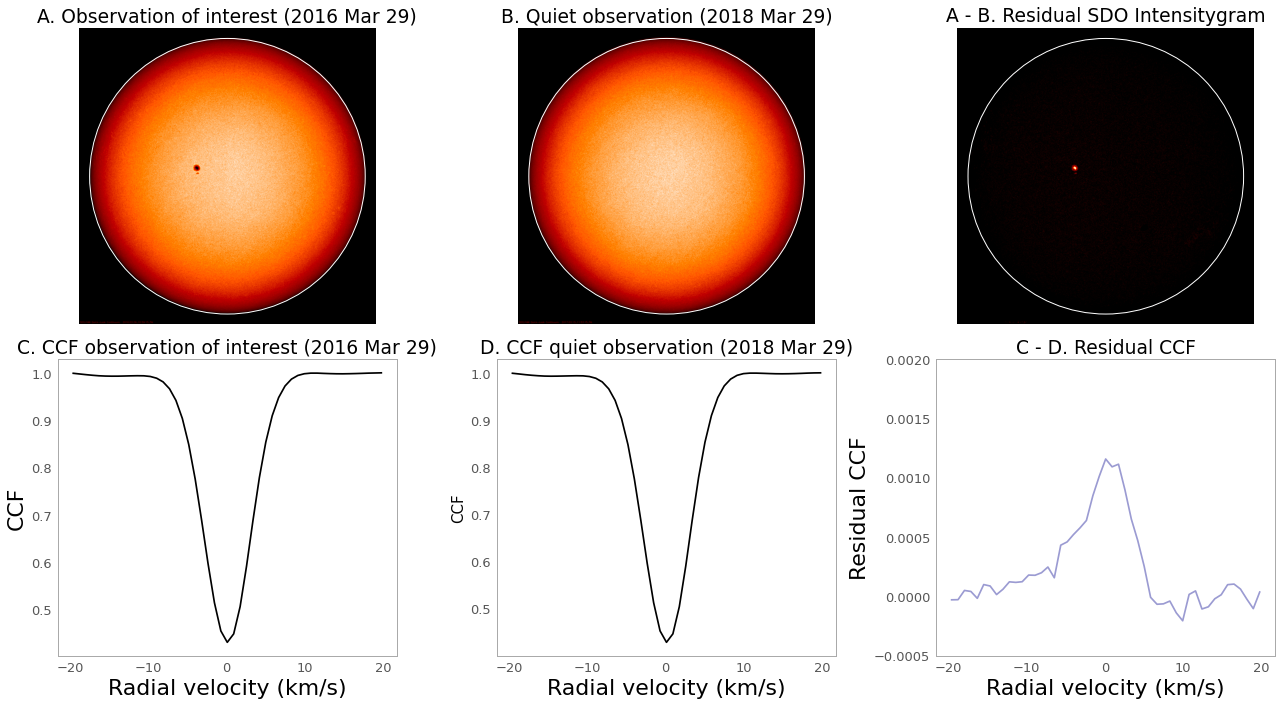}
    \caption{Residual CCF ($\Delta$CCFs) construction: top row SDO/HMI intensitygrams; bottom row CCFs from HARPS-N Solar Telescope observations. The left column (panels A, C) are from a period of relatively large solar activity while the middle column (panels B, D) are from observations of the quiet sun. To highlight differences in shape between CCFs, in the rightmost column, we subtract a quiet observation (B/D) from the observation of interest (A/C).
    }
    \label{fig:oneresidual}
\end{figure*}

\begin{figure*}
    \epsscale{1.2}
    \plotone{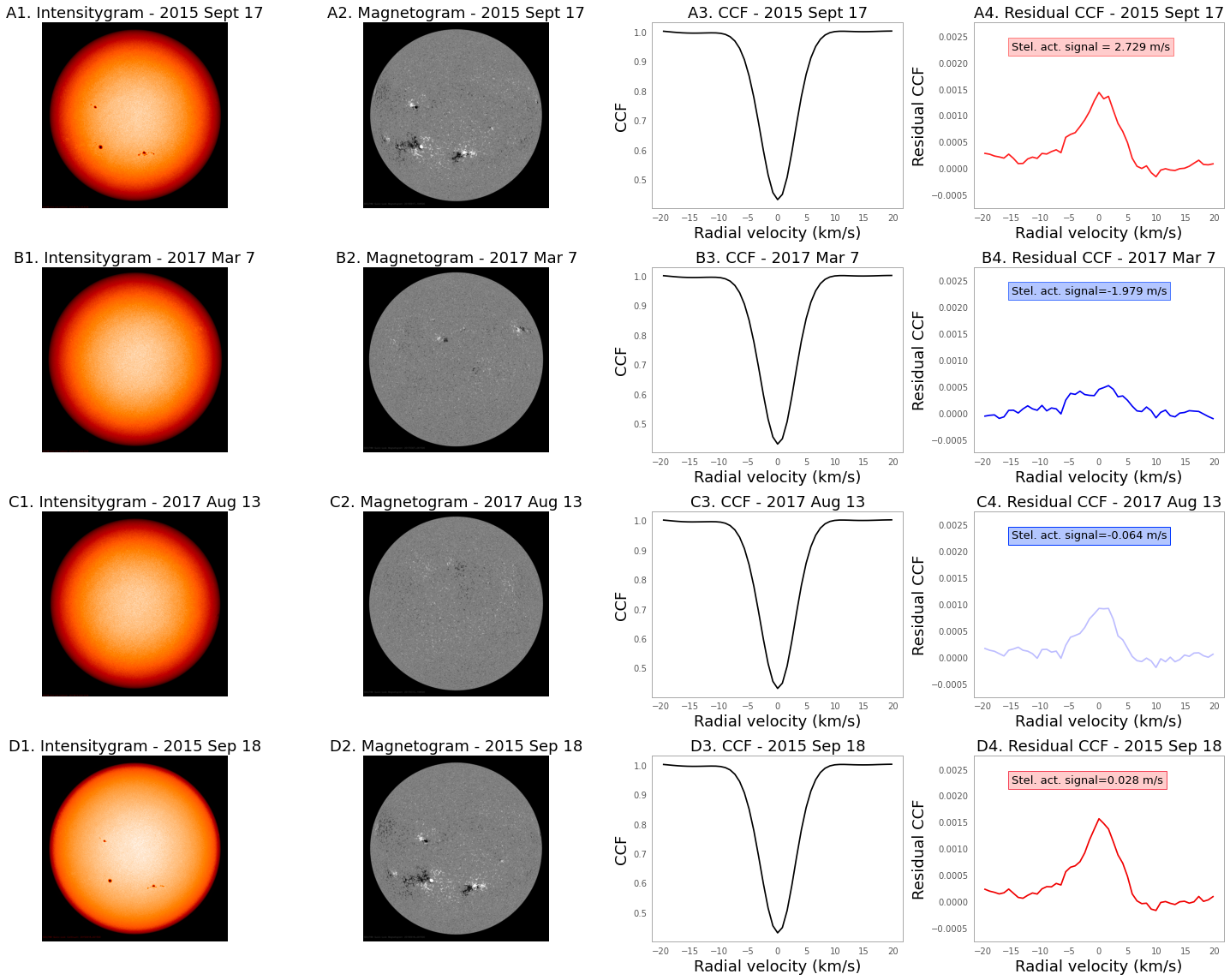}
    \caption{
    Comparison of Residual CCF ($\Delta$CCFs) and SDO Observations: Column 1 (A1, B1, C1, D1) are SDO/HMI intensitygrams; Column 2 (A2,...,D2) are SDO/HMI magnetograms; Column 3 (A3,...,D3) are CCFs from HARPS-N Solar Telescope observations; Column 4 (A4,...,D4) are Residual CCFs from HARPS-N where the corresponding RVs are indicated by their color (red = redshifted, blue = blueshifted). Each of the rows are from a different day of observations and subsequently display distinct surface inhomogeneities that correspond to the residual CCF line shape (A4, B4, C4, D4).}
    \label{fig:exampleresiduals}
\end{figure*}

\section{Data}\label{observation_descrip} %How do we go from photons or random number generators to CCFs? 

ML methods require data on which to learn. We trained ML models on two datasets: one set of simulated stellar spectra from the SOAP 2.0 software \citep{2014ApJ...796..132D} and one set of real RV observations of the Sun from the HARPS-N solar telescope \citep{dumusque2015, 2016SPIE.9912E..6ZP, 2019MNRAS.487.1082C}. 

\subsection{SOAP simulations}\label{soap_data}

We first explored the problem of predicting stellar activity variations with a simulated dataset. We produced this dataset using SOAP 2.0, a software package which estimates photometric and RV variations induced by sunspots and faculae \citep{2014ApJ...796..132D, 2012A&A...545A.109B}. SOAP simulates a star by dividing the visible hemisphere into a grid and injecting in each created cell an observed solar cross-correlation function (CCF; obtained by cross-correlating a solar spectrum by a binary mask, as it is done when reducing HARPS-N high-resolution spectra). The CCFs are shifted to account for the star's rotation velocity at each point on the star's surface. In certain user-specified locations, SOAP 2.0 modifies the local CCFs to mimic active regions, like dark spots or faculue. Finally, SOAP 2.0 sums the CCFs over the entire visible hemisphere of the star, convolves the resulting CCF with a simulated instrumental line profile, and fits the result with a Gaussian function to derive the RV due to the activity signal. 

%To train a machine learning model to regress out stellar activity signals, a large randomly distributed data set was required. Thus, we developed a Monte Carlo simulation as a training set using SOAP 2.0, software which estimates photometric and RV variations induced by sunspots and faculue \citep{2014ApJ...796..132D}. 

We modified SOAP 2.0 to produce a large set of simulated observations with varying parameters chosen with a Monte Carlo technique. We generated 20,000 random starspot/faculae configurations and used SOAP 2.0 to produce a simulated CCF and activity RV measurement for each configuration. \textit{Table \ref{tab:soapparameters}}  lists the range of values that each of the stellar parameters spans.

\subsection{HARPS-N Solar Telescope}\label{harps_data}

The HARPS-N Solar dataset consists of \bedit{528} days of solar observations (See \textit{Section \ref{harps_input}} for details) from the HARPS-N Solar Telescope spanning three years (July 2015 - July 2018). Commissioned at the Telescopio Nazionale Galileo (TNG), the HARPS-N spectrograph is a vacuum-enclosed cross-dispersed echelle spectrograph that has temperature and pressure stabilization \citep{2012SPIE.8446E..1VC}. HARPS-N spans the wavelength range from 383 to 693 nm and has a resolving power of $\lambda/\Delta\lambda=$ 115,000.
During the daytime, a custom-built solar telescope connected to HARPS-N continuously observes the Sun with 5-minute integration times designed to mitigate the short-term variability caused by solar 5-minute p-mode oscillations.  The solar telescope and control system are further described by \citet{2016SPIE.9912E..6ZP}.

The solar data are reduced using the same HARPS-N Data Reduction System (DRS) as used for nighttime stellar observations. By taking calibration exposures at the end of each day of solar observations, we acquire order-by-order information on the locations of the echelle orders and the wavelength calibration scale. The instrumental drift is monitored by taking exposures of light passed through a stabilized Fabry-Perot cavity concurrently with the solar exposures. After applying optimal extraction procedures \citep{1986PASP...98..609H,1989PASP..101.1032M}, the data are calibrated in wavelength such that we can obtain a 1D background subtracted spectrum in each echelle order. Lastly, the data are cross-correlated with a digital mask \citep{1996A&AS..119..373B, 2002A&A...388..632P} derived from solar observations and corrected for instrumental drift based on the Fabry-Perot exposures. The resulting CCF is used for our input representation to the ML method. Finally, the DRS extracts the RV of each observation by fitting the CCF with a Gaussian function. A Gaussian function is simple and symmetric. Therefore, it is unable to model the small perturbations to CCF shapes induced by stellar activity. So these RVs include both center of mass RV shifts and stellar activity signals. 

We note that the full three-year dataset of HARPS-N used in our paper was recently released to the public, as described by \citet{dumusque2020}. The data products released by \citet{dumusque2020}\footnote{\url{https://dace.unige.ch/sun/?}} were reduced with a new extraction pipeline originally built for the newly commissioned ESPRESSO instrument. This new ESPRESSO pipeline analysis resulted in more precise RV measurements than the original HARPS-N DRS reductions \citep{2019MNRAS.487.1082C} \bedit{and were thus used in this analysis.}

%but we observed that there were additional changes to the shape of the CCFs that hurt the performance of our ML methods. We therefore use the original HARPS-N DRS data products in our analysis and leave an analysis of the new CCFs for future publications.

\section{Preparing the input representations}\label{inputrepresentation}% How do we go from CCFs to something that can go straight into Tensorflow? 

For our ML models, we cannot use the data directly from the telescope or simulations. Instead, we have to pre-process these data products into a uniform format that makes capturing the features in the data easier for ML models. We outline the steps we took to prepare the input representation for the ML models both for the simulated (\textit{Section \ref{soapinput}}) and HARPS-N Solar Telescope Data (\textit{Section \ref{harps_input}}). 

We design the input representations to pose the problem to our ML models of predicting the activity signal, not the actual center-of-mass velocity of the star. Essentially, we want the ML model to predict \textit{the difference between a Gaussian fit to the CCF and the true velocity shift}. With these predictions in hand, we can easily subtract them from the input RVs to produce a corrected RV time series.

%To frame the problem, we shift every CCF such that the velocity predicted by the Gaussian fit to the CCF is at 0. 

Intuitively, RV signals due to planets cause translational shifts on the CCF, but do not result in shape changes of the CCF. In contrast, stellar activity does not result in translational changes and only causes shape changes. Thus, we want the measured RV of our simulated CCF to be shifted to 0 so that the ML methods become primarily sensitive to detecting these shape changes, not translations.

%It may also make sense to state this in the last paragraph of the intro. 
%\vspace{0.4in}
\subsection{SOAP Input Representation}\label{soapinput}

Given the CCFs generated by SOAP 2.0 and the measured RV signals (due to the simulated active regions on the star), we apply the following pre-processing steps before sending the CCFs (without any timing information) into our ML models: 

\begin{enumerate}
\item First we take the simulated CCF from SOAP 2.0, and shift it by the velocity measured by the SOAP's Gaussian fit to the CCF. We do this by creating an $x'$ axis that is shifted by $-\Delta RV$, where $\Delta RV$ is the stellar activity shift measured by SOAP and interpolating the CCF values from the $x'$ axis to the original $x$ axis by using \texttt{scipy.interpolate.interp1d()} to perform a cubic interpolation. We tested multiple interpolation methods (linear, nearest, cubic) and found that the cubic method was optimal. We also confirmed that any systematics introduced by interpolation were smaller than the changes due to stellar activity. (\textit{Note:} In the presence of keplerian shifts this procedure would need to be modified as described in \textit{Section \ref{futurework}}). \bedit{Shifting the CCF to the velocity measured by the SOAP Gaussian fit purposefully leaves a small translational shift between the true stellar radial velocity and 0; this shift is exactly what we wish to train the model to predict.}

\item We then calculate a differential $\Delta$CCF by subtracting a template CCF generated by SOAP with no active regions (also shifted as described in the previous step). \bedit{We note that choosing a template CCF with no active regions or another random template CCF does not affect the overall analysis results, but} this particular choice of $\Delta$CCF highlights the changes to the shape of the CCF introduced by the active regions. 

\item Lastly, we normalize the inputs to the ML methods. In particular, since each input is an array comprising the $\Delta$CCF, we calculate the median and standard deviation of each point in the CCF over the entire simulated dataset, and normalize by subtracting the median and dividing by the standard deviation. This helps the optimization process by making the scale of variations of each input parameter roughly equal. 

\item Each input into the ML methods is only this normalized $\Delta$CCF without any timing information. The neural network is then trained to predict the stellar activity signals only based on shape differences between normalized $\Delta$CCFs from different observations.
\end{enumerate}

%\textcolor{red}{Maybe make this an itemized list}

%We took the simulated CCF from SOAP, shift it, subtracted an activity free CCF, and normalized the input parameters by subtracting the median and dividing by the standard deviation. 

%This stellar activity RV signal is directly output by SOAP2.0, but for the CCF we have to make some adjustments.  Thus, we shift the x-axis by the stellar activity RV signal from SOAP2.0 and then use interpolation to shift the CCF to zero. 

%The procedure by which we shift the CCF is: 

%\begin{enumerate}
%    \item  We create an $x'$ axis that is shifted by $-\Delta RV$, where $\Delta RV$ is the stellar activity shift we want to introduce
%    \item We interpolate the CCF values from the $x'$ axis to the original x axis (using scipy interp1d). 
%\end{enumerate}

%Lastly, we scale the $\Delta$CCF by subtracting the median and dividing by the standard deviation. Finally, this scaled $\Delta$CCF serves as the input representation and the stellar activity RV signal serves as the output for the machine learning method.

\begin{figure}[ht!]
\centering
	\includegraphics[width=3.5in]{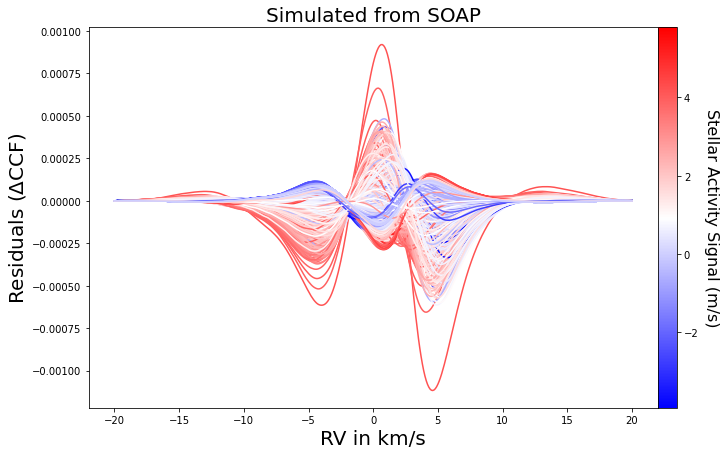}
    \caption{Simulated $\Delta$CCFs from our Monte Carlo training set generated by SOAP 2.0 before normalization. The different shapes are the result of different activity configurations on the star. The measured RV activity signal for each $\Delta$CCF is indicated by its color (red = redshifted, blue = blueshifted). }
    \label{fig:simresiduals}
\end{figure}

\subsection{HARPS-N Input Representation}\label{harps_input}

Our pre-processing for the HARPS-N data is nearly identical to our pre-processing for the SOAP data, with only two additional steps. Our procedure is as follows:

\begin{enumerate}
    \item Create a daily average of CCFs. During the day, HARPS-N takes repeated 5 minute-long exposures of the Sun. We average all of these exposures together to obtain a daily averaged CCF and RV measurement. To do this, we follow \citet{CollierCameron2021}. In short, we perform a signal-to-noise weighted average of the CCFs and RVs. We exclude individual observations where the probability of being cloud-free (as calculated by \citealt{2019MNRAS.487.1082C}) is greater than 99\% and where the expected differential extinction\footnote{For observations of an extended source such as the Sun, we must consider how the gradient in atmospheric extinction across the star's disc results in asymmetries in the CCF. This phenomenon is often referred to as differential extinction and this gradient has different effects on blue versus red components \citep{1972BAICz..23...75R}. As the solar disc rotates, the CCF is rotationally broadened and this systematic signal results in asymmetries of the CCF which can even be mistaken for solar oscillations in some cases \citep{1979A&A....77..351G,1980A&A....88..317S}.} correction is less than 10 \cms. 
    \item Remove signals from Solar System planets. The raw RVs measured by the DRS consist of both the radial motion induced by the solar system planets and stellar activity signals. The planetary signal is dominated by a sinusoidal signal with a semiamplitude of 12 \ms\ and a period of $\sim$ 13 months, which is the synodic period of Jupiter observed from Earth. To remove the planetary signals, we transform both the RVs and the CCFs from the barycentric to the heliocentric reference frame using the JPL Horizons ephermeris \citep{1996DPS....28.2504G}. For the RVs, we simply subtract the Sun's barycentric motion in the direction of the TNG to derive the heliocentric RV. For the CCFs, we perform the shift with the same method as we used to shift the SOAP CCFs, by creating a shifted $x'$ axis and interpolating back onto the $x$ axis. The resulting velocities and CCFs contain only stellar activity shifts (and instrumental systematics), in analogy to the simulated CCFs produced by SOAP 2.0. 
    \item After removing the Solar System planet signals, we shift the CCF so that the velocity measured from the HARPS-N DRS Gaussian fit is 0. This step is identical to Step 1 from our SOAP preprocessing in \textit{Section \ref{soapinput}}. As a reminder, we take this step because we know that RV signals from planets cause translational shifts, not shape changes. In contrast, stellar activity results in shape changes and no translational shifts. Thus, we shift the CCF and RV to 0 so that the ML methods become primarily sensitive to detecting these shape changes, not translations.
    \item We calculate the differential $\Delta$CCF by subtracting a reference HARPS-N observation taken when the Sun had few magnetic features on its visible hemisphere, as determined by visual inspection of images from the Solar Dynamics Observatory (SDO) Helioseismic and Magnetic Imager (HMI). The observation we used as a quiet reference is from March 29, 2016 (See \textit{Figure \ref{fig:oneresidual}}). \bedit{Although choosing a random other template CCF yields the same overall results, choosing a template with few magnetic features on its visible hemisphere allows us to visualize CCF shape changes as a function of activity more clearly.} This step is analogous to Step 2 from our SOAP preprocessing in \textit{Section \ref{soapinput}}. 
    \item We normalize the input features in exactly the same way as described in Step 3 from our SOAP preprocessing in \textit{Section \ref{soapinput}}.
    \item Each input into the ML methods is only this normalized $\Delta$CCF without any timing information. The neural network is then trained to predict the stellar activity signals only based on shape differences between normalized $\Delta$CCFs from different observations.
\end{enumerate}

\begin{figure*}
%\begin{interactive}{animation}{Figure4.mp4}
%\plotone{Figure4.png}
%\end{interactive}
\includegraphics[width=1\linewidth]{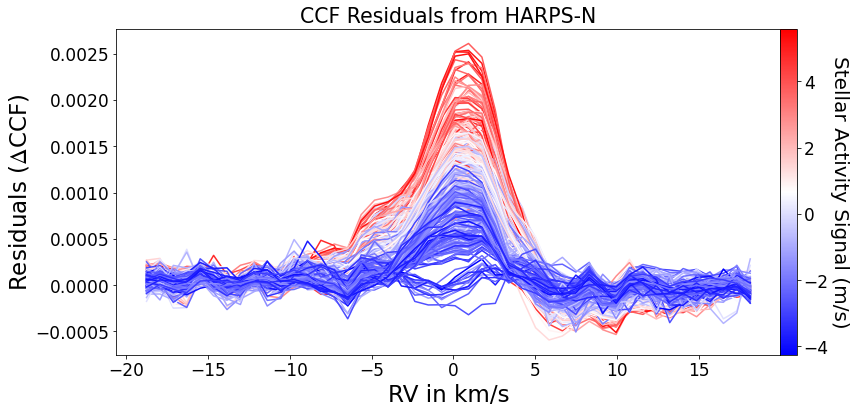}
\caption{HARPS-N $\Delta$CCFs before normalization
    \textit{--} Computed residual CCFs ($\Delta$CCFs) by subtracting the mean CCF, highlighting differences in features between CCFs. For training the model, $\Delta$CCF is the input and the RV from stellar activity is the output. The RV is indicated by its color (red = redshifted, blue = blueshifted). An animated version of this figure is available online$^4$ has a duration of 53 seconds. In this animated version, the CCF residuals are layered on top of each other in the order with which they were observed. At the end of the animation, we reach the same final frame as is displayed in this static version of the Figure.}
    \small\textsuperscript{4} \url{https://github.com/zdebeurs/exoplanet-ml/tree/master/exoplanet-ml/rv_net}
    \label{fig:residals}
\end{figure*}

% old version of Figure 4 
\iffalse 
\begin{figure*}[ht!]
\centering
	\includegraphics[width=6.5in]{Figure4.png}
    \caption{HARPS-N $\Delta$CCFs before normalization
    \textit{--} Computed residual CCFs ($\Delta$CCFs) by subtracting the mean CCF, highlighting differences in features between CCFs. For training the model, $\Delta$CCF is the input and the RV from stellar activity is the output. The RV is indicated by its color (red = redshifted, blue = blueshifted). An animated version of this figure is available online$^4$ \bedit{has a duration of 53 seconds. In this animated version, the CCF residuals are layered on top of each other in the order with which they were observed. At the end of the animation, we reach the same final frame as is displayed in this static version of the Figure.}}
    \small\textsuperscript{4} \url{https://github.com/zdebeurs/exoplanet-ml/tree/master/exoplanet-ml/rv_net}
    \label{fig:residals}
\end{figure*}
\fi

\subsection{Visualizing the inputs}

The result of these processing steps is a residual $\Delta$CCF for each observation in our dataset. Here, we hope to give an intuitive understanding of what these residual $\Delta$CCFs represent and how they convey information about stellar activity.  In \textit{Figure \ref{fig:oneresidual}}, we illustrate how we subtract a quiet observation (panels B and D) from the observation of interest (panels A and C) to calculate the residual CCF ((A-B) and (C-D)). We note that the residual $\Delta$CCFs shown here have not yet been scaled by subtracting the median and dividing by the standard deviation. \textit{Figure \ref{fig:exampleresiduals}} shows several example $\Delta$CCFs taken on different dates, illustrating how different activity patterns change the $\Delta$CCFs. 

In \textit{Figures \ref{fig:simresiduals} and \ref{fig:residals}} (animated version available online\footnote{\url{https://github.com/zdebeurs/exoplanet-ml/tree/master/exoplanet-ml/rv_net}}), we illustrate the differences in shape of the $\Delta$CCFs for different RVs induced by sunspots and faculae over our entire dataset. Each $\Delta$CCF is color-coded to show the measured RV induced by stellar activity. Clear patterns emerge in these plots where similar $\Delta$CCF shapes tend to have similar measured RVs. These are the patterns our ML methods will use to predict stellar activity signals from the $\Delta$CCFs which we can use to correct stellar activity.  % Changes in the $\Delta$CCF shape serve as the input features based on which the machine learning method has to produce the output radial velocity induced by stellar activity. 

\section{Creating Training, Validation, and Test Sets}\label{train_val_test}
In ML, datasets are commonly randomly separated into a training, validation, and testing set. The model is initially fit on the training set, a set of examples used to fit the parameters of the model. Next, the validation set provides a measure of predictive accuracy and model fit. The validation set consists of examples that the model has not seen in the training set and allows for optimization of the architecture and hyperparameters. Lastly, after the model architecture and hyperparameters are finalized, the test set is used as one last objective test of the model accuracy and fit.

In our work, we divided our two datasets (from SOAP 2.0 and the HARPS-N Solar Telescope) into separate groups for training, validation, and testing.  Since the simulated SOAP 2.0 training set is sufficiently large, we divided the dataset into training (80$\%$ of the data), validation (10$\%$), and testing sets (10$\%$). %To train our ML models, we simulated a Monte Carlo training set of 20,000 observations and used 629 days of radial velocity observations from the HARPS-N Solar Telescope.

However, our smaller dataset from the HARPS-N Solar Telescope required a different approach. Instead, we created a cross-validation set (80$\%$ of the dataset), a validation set (10$\%$; which was trained on the full cross-validation set), and a testing set (10$\%$). We then use k-fold cross-validation to provide as many tests with the available data to optimize the architecture and hyperparameters. In k-fold cross-validation, the cross-validation dataset is divided into k subsets. For each round of validation, one of the k subsets is treated as a holdout sample and the model is trained on the other k-1 subsets. In this way, k-fold cross-validation significantly decreases bias (i.e. overestimate of model performance) as we are using the majority of the data for fitting. The exact choice of k represents a trade-off between bias and variance (i.e. performance changes significantly based on data chosen to train the model). A higher value of k may decrease the variance but can also increase the bias. We divided our cross-validation set into 10 folds as k=10 has been shown empirically to yield test error estimates that minimize both bias and variance \citep{james2013introduction}. We optimized the architectures by assessing the performance on both the k-fold cross-validation set and the held-out 10\% validation set (which we trained on the full cross-validation set). To estimate how well the final model generalizes, we evaluated our best models' performances on the test set (10 $\%$ of the dataset), which consists only of examples that were not used in the cross-validation or validation.

\section{Neural Network Architectures}\label{architectures}
%\subsection{Neural Network Architecturse}

\begin{figure*}
    \epsscale{1.05}
    \plotone{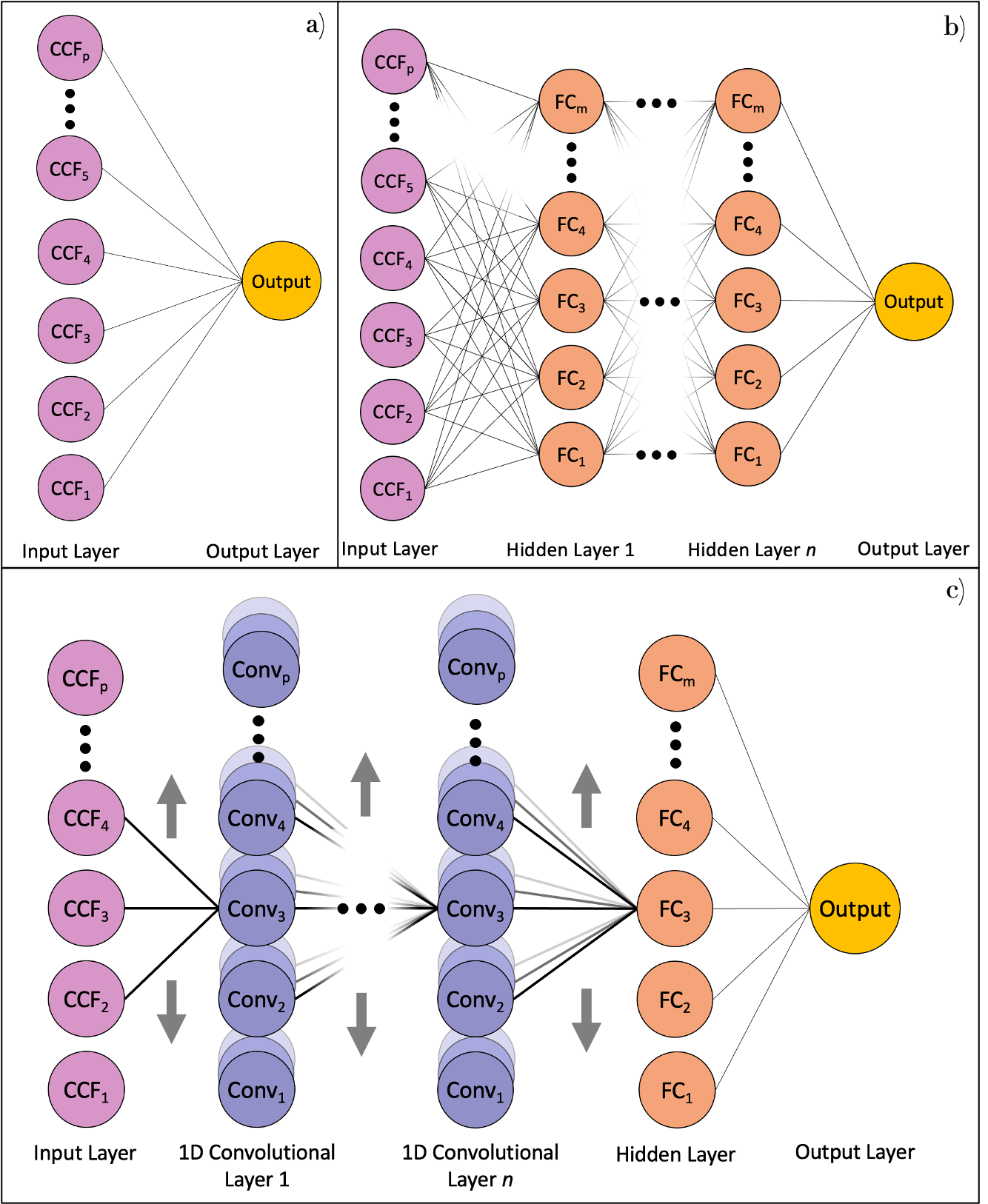}
    \caption{Three ML Architectures Visualized - In all three architectures, the vector of all parameters in the model $\bm{p}$ matches the input dimensions and $\bm{p}$  = 401 for SOAP2.0 data and $\bm{p}$  = \bedit{46} for HARPS-N data. a) Linear Architecture. This architecture is equivalent to a linear regression model and has zero hidden layers. b) Fully Connected Neural Network (FC NN). Every connection corresponds to a multiplicative weight parameter learned by the model. The CCF inputs are fed into the first layer, the hidden layers represent a hierarchy of learned features, and the output layer generates predictions. For the final FC model, the number of dense units m = 100 for SOAP2.0 data and m = 200 HARPS-N data. The number of hidden layers n = 1 for SOAP2.0 data and n = \bedit{8} for HARPS-N data. c) Convolutional Neural Network (CNN). The convolutional layer takes the discrete cross-correlation of the vector in its input layer with the kernel vectors that the model learns. The sparse connections compared to the FC model allow CNNs to learn local features and exploit spatial structure in the data. The stack of nodes going into the page within each convolutional layer represent the different filters. For the final CNN model, the number of 1D convolutional layers n = 6 and m = 1000 for the SOAP2.0 data and \bedit{n = 1 and m = 500 for the HARPS-N data}. In all three models, the ellipses ($\ldots$ and $\vdots$) represent additional nodes and layers which we have omitted for visual clarity. }
    \label{fig:sllarchitectures}
\end{figure*}

\begin{figure}[ht!]
\centering
	\includegraphics[width=1.5in]{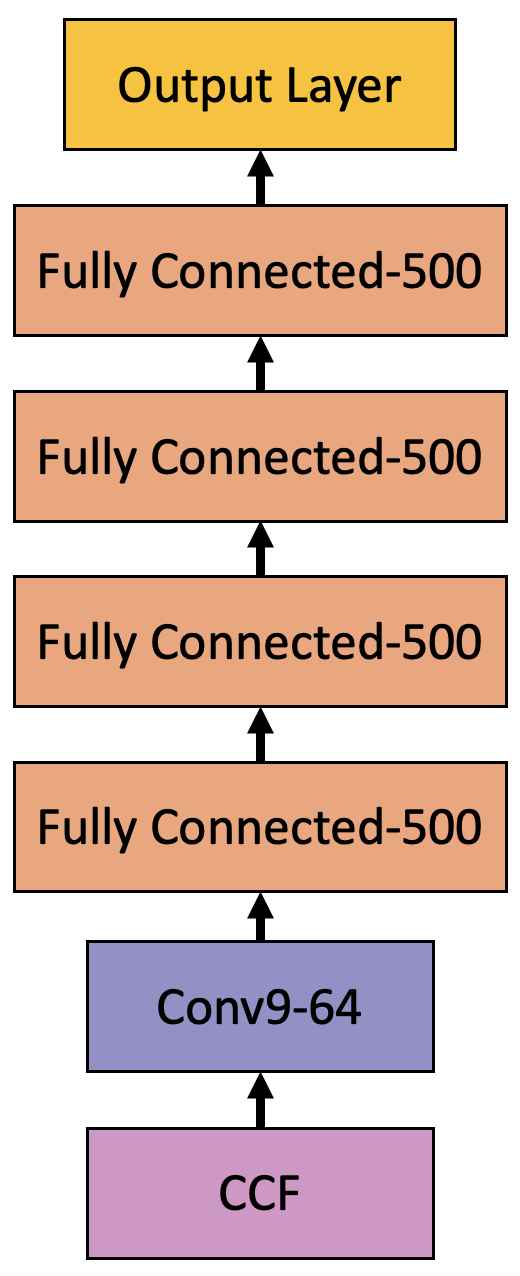}
    \caption{The architecture of our best performing neural network model
    \textit{--} Convolutional layers are denoted conv$\langle$kernel size$\rangle$-$\langle$number of feature maps$\rangle$, fully connected layers are denoted Fully Connected - $\langle$number of units$\rangle$.}
    \label{fig:cnnarchitecture}
\end{figure}

To learn to predict stellar activity RV signals from differences in shape of the $\Delta$CCF, we trained three different ML architectures: a linear regression model, a fully connected neural network, and a convolutional neural network.

\subsection{Linear Architecture}

The most basic model we trained is a linear regression model, which is equivalent to a zero hidden layer neural network. As illustrated in \textit{Figure \ref{fig:sllarchitectures}a}, the linear architecture takes a vector $\bm{x}\in 	\mathbb{R}^{n}$ as an input where $\mathbb{R}^{n}$ is a real coordinate space of dimension $n$ (where $n = $ dimension of the input data). \bedit{The input vector $\bm{x}$ represents the rescaled CCF residuals}. After taking the vector $\bm{x}$, the linear architecture predicts the value of a scalar $y$ as the output\bedit{, which is the predicted stellar activity RVs}. Then the predicted value of $y$ will be

\begin{equation}
    \hat{y} = \bm{w}^{\top} \bm{x} + b
\end{equation}

where $\bm{w}\in \mathbb{R}^{n}$ are the weights that determine how each feature affects the prediction, and $b$ is a bias term. $\bm{w}$ and $b$ are the trainable parameters of the model.

Although a convenient choice due to its simplicity, a linear architecture makes the strong assumption that a linear relationship exists between the points in the CCF and the RV activity signal. For simulated ideal data this assumption may not pose a problem, but for more complex real data this assumption can break down (See Section \ref{results}).

\subsection{Fully Connected Architecture} 
\textit{Figure \ref{fig:sllarchitectures}b} shows a fully connected neural network (FC NN; also sometimes referred to as a multilayer perceptron or feed-forward neural network). Each layer consists of scalar-valued units called \textit{neurons} where the outputs from one layer of neurons are the inputs for the next layer. The function that produces outputs based on the inputs is called an activation function. This activation function  $\phi$ produces a new representation of  $\bm{w}^{\top} \bm{x} + b$ through a nonlinear transformation; its output $\phi(\bm{w}^{\top} \bm{x} + b)$ can be thought of as a set of features describing $\bm{x}$.

The values of the first and last layers comprise the inputs and outputs of the network. However, the values (activations) of the intermediate layers are not directly observed and are therefore referred to as hidden layers. The hidden and output layer activations are defined by

\begin{equation}
    \bm{a}_n = \phi (\bm{W}_{n} \bm{a}_{n-1} + \bm{b}_n)
\end{equation}

where $n$ is the layer number, $\bm{a}_n$ is a $i_n$-long vector of activations in layer $n$, $\bm{W}_n$ is an $i_n \times i_{n-1}$ matrix of learned weights, $\bm{b}_n$ is a $i_n$-long vector of learned bias parameters, and $\phi$ specifies an activation function that computes the hidden layer values. 

In FC NNs, the most common activation function is the rectified linear unit \citep[ReLu;][]{jarrett2009best, nair2010rectified, glorot2011deep}, defined by the element-wise activation $\phi({x}) = max\{0,{x}\}$. The element-wise activation $\phi(x)$ applies a nonlinear transformation where values of $x < 0$ are mapped to zero and others remain equal to $x$. In this way, rectified linear units are nearly linear and preserve many of the properties that make linear models easy to optimize with gradient-based methods \citep{Goodfellow-et-al-2016}. In our neural network layers, we used ReLu as our activation functions. 

\subsection{Convolutional Architecture}

FC NNs use matrix multiplication where the matrix has a separate parameter for the interaction between each input unit and every output unit. Every input interacts with every output, causing FC NNs to be ``agnostic'' to spatial structure present in the data. For example, they treat adjacent data points exactly the same as data points that are far apart, making it inefficient to learn local features (e.g. edges and shapes) that may appear in different locations. In contrast, convolutional neural networks (CNNs) have only local (sparse) interactions, which force them to learn local features across the entire input and exploit the spatial structure (\textit{Figure \ref{fig:sllarchitectures}c}). Rather than learning local features for every single input-output interaction, they only have to learn these features once. This reduces the number of parameters that the model needs to learn and decreases the number of computational operations required to predict the output.

The 1D convolutional layers depicted in \textit{Figure \ref{fig:sllarchitectures}c} require an input stack of $K$ vectors $\bm{a}_{n-1}^{(k)} (k=1,2,...K)$ of length $i_{n-1}$. Each convolutional layer outputs a stack of L vectors $\bm{a}_{n}^{(l)} (l=1,2,...L)$ of length $i_{n}$. The transformation that takes the stack of $K$ input vectors and produces the $l$th output vector is called a feature map defined by the operation

\begin{equation}
    \bm{a}_n^{(l)} = \phi \bigg( \sum_{k=1}^{K} \bm{w}_n^{(k,l)} * \bm{a}_{n-1}^{(k)} + \bm{b}_{n}^{(l)} \bigg)
\end{equation}

where $*$ is the discrete cross-correlation function (often referred to as a ``convolution'' in the ML literature), $\bm{w}_n^{(k,l)}$ is a $m_n$-long vector of learned parameters called the convolution kernel or filter, $\bm{b}_n$ a $i_n$-long vector of learned bias parameters, and $\phi$ specifies the activation function. By making the kernel size small ($m_n \sim 3-7$), the feature map becomes sensitive to local features along its input.

Intuitively, we would expect the CNN to perform best because we are trying to identify shapes and relationships between adjacent points in our $\Delta$CCF to produce the final stellar activity output.

Commonly, CNNs use both convolutional layers and pooling layers. Pooling layers typically replace the output of the neural net at a certain location with a summary metric (e.g., maximum) of the nearby outputs. Pooling generally helps to make the representation translationally invariant. However, in early iterations of our models, we found that adding pooling layers negatively affected our performance so we do not use any pooling layers in our CNN models.

\bedit{After the convolutional layers in our CNN model, we finally include one (or more) fully connected layer(s). The last fully connected layer then produces the final output.}

\section{Neural Network Training}\label{training_proc}

\subsection{Training algorithm}

Neural networks are trained to minimize a loss function, which quantifies the difference between the predictions and the true labels in the training set. For regression problems, the mean squared error (MSE) is the standard loss function and is defined by the equation
\begin{equation}\label{loss}
    L(\hat{y}_i, y_i|\bm{p}) = MSE =  \frac{1}{M} \sum_{i=1}^{M} ( \hat{y}_i-y_i)^{2}
\end{equation}
where $\bm{p}$ is the vector of all parameters in the model, $y_1, y_2, ..., y_M$ are the true labels of all examples in the training set, and $\hat{y}_1, \hat{y}_2, ..., \hat{y}_M$ are the model's predicted outputs given $\bm{p}$. \bedit{The vector $\bm{p}$  consists of all the free parameters of the given architecture to be learned during training. For the linear architecture, this is simply the vector of weights $\bm{\omega}^{\top}$ and the bias term $b$. For the FC NN, this corresponds to the elements of all weight matrices $\bm{W}_n$ and the bias vectors $\bm{b}_n$. For the CNN, the parameters are the elements of all convolutional kernels (or filters) $\bm{w}_n^{(k,l)}$, bias vectors $\bm{b}_n$, and the weight matrix and bias matrix of the final fully connected layer.}

The most popular neural network training algorithms use gradient descent to find the parameters $\bm{p}$ that minimize the loss function. These algorithms calculate how the parameters of the model can be changed to decrease the loss function by computing the gradient of the loss function with respect to the parameters. The model's parameters start at random values and are iteratively updated by descending along the gradient until the desirable minimum of the loss function is achieved. The step size is set by the learning rate, which is a hyperparameter that requires tuning during (cross-)validation to achieve optimal performance.

Computing the exact gradient of the loss function is unnecessary and computationally inefficient as it requires iterating over the entire training set. Rather, each gradient step approximates the true gradient by taking a random batch (i.e. subset) of the training set. The algorithm is then called a stochastic gradient descent (SGD) algorithm. The batch size is typically determined by the available computational resources. We kept the batch size constant at 300 -- \citet{shallue2018measuring} demonstrated that the performance should be the same at any batch size, provided the other hyperparameters are well-tuned.

In practice, neural networks are often trained using variants of the basic SGD algorithm. For our FC and CNN neural networks, we used SGD with momentum \citep{polyak1964some} with the momentum parameter fixed at 0.9.

\subsection{Overfitting and Regularization}
The fundamental challenge in ML is that algorithms have to perform well on \textit{novel, previously unseen} inputs - not just the data on which the model was trained \citep{Goodfellow-et-al-2016}. The ability of a model to perform well on previously unseen inputs is called generalization, and can be estimated from its performance on a test set comprised of data not used during training.

Overall, we can summarize the performance of a ML method by its ability to minimize both 1) the training error and 2) the gap between the training and test error. These two abilities correspond to the problems of underfitting and overfitting on training data, respectively. Techniques that seek to reduce overfitting, and therefore improve generalization, are known as regularization methods.

A common approach to regularization is to limit the complexity of the model by constraining the values of its parameters $\bm{p}$. Two such methods are $L_{2}$ regularization and weight decay regularization. $L_{2}$ regularization adds a penalty term to the loss function proportional to the squared $L_{2}$ norm of the parameter vector $\bm{p}$,
\begin{equation}
    L_{reg} = MSE + \frac{\color{black}{\alpha}}{2} ||\bm{p}||_{2}^{2},
\end{equation}
where \bedit{$\alpha$} is the strength of the regularization\footnote{\bedit{Sometimes, this regularization strength parameter is referred to as $\lambda$ in the literature. We choose $\alpha$ here for consistency with our description of our our linear model regularization procedure in Section \ref{lin_mod_impl}.}} and the  $L_{2}$ norm $||\bm{p}||_{2}$ is defined as
\begin{equation}
    ||\bm{p}||_{2} =  \left(\sum_{i=1}^{N} |\bm{p}_i|^2 \right)^{1/2} = \sqrt{\bm{p}_1^2+\bm{p}_2^2+ .... + \bm{p}_N^2}
\end{equation}
Weight decay regularization shrinks the parameter vector by a constant factor on each iteration,
\begin{equation}
    \bm{p}_{i+1} = (1-{\color{black}{\alpha}}) \bm{p}_{i}  + (\Delta \bm{p}_i)_\text{opt},
\end{equation}
where $(\Delta \bm{p}_i)_\text{opt}$ is the change to the parameter vector computed by the optimization algorithm at iteration $i$. Both of these techniques encourage smaller values of the parameters of $\bm{p}$. In fact, $L_{2}$ regularization and weight decay are equivalent when using the basic SGD training algorithm. However, they are not equivalent for all variants of SGD, in particular for SGD with momentum, which we used for our FC and CNN neural networks. In those cases, we chose weight decay because it empirically performs better for neural networks \citep{2017arXiv171105101L}.

Larger values of the weight decay parameter $\alpha$ discourage overfitting, but can also cause underfitting.  We optimized $\alpha$ by exploring the parameter space during validation for the SOAP 2.0 simulated data and during cross-validation for the HARPS-N data for both our FC and CNN models.

\subsection{Implementation and Training Procedure}

For each of the model architectures, we tune the hyperparameters unique to the architecture. In contrast with parameters that are learned during the training process, hyperparameters are parameters whose value is used to control the learning process. The hyperparameters can significantly affect model performance.

\begin{figure}[ht!]
\centering
	\includegraphics[width=0.5\textwidth]{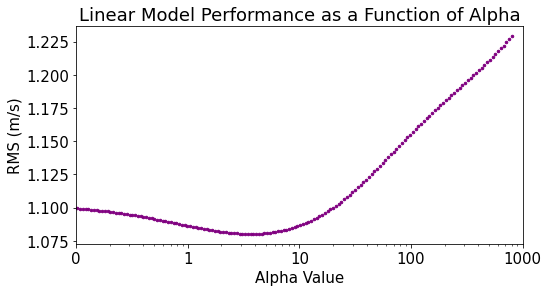}
    \caption{\bedit{Linear model hyperparameter $\alpha$ Optimization
    \textit{--} The Root-Mean Square (RMS) Error for the cross-validation set as a function of the value of alpha $\alpha$ is listed. The value of $\alpha$ with the lowest corresponding RMS is listed in Table \ref{harpsnvalcrossvaltablelinear}.}}
    \label{fig:alphalinear}
\end{figure}

\begin{figure*}[ht!]
\centering
	\includegraphics[width=1\textwidth]{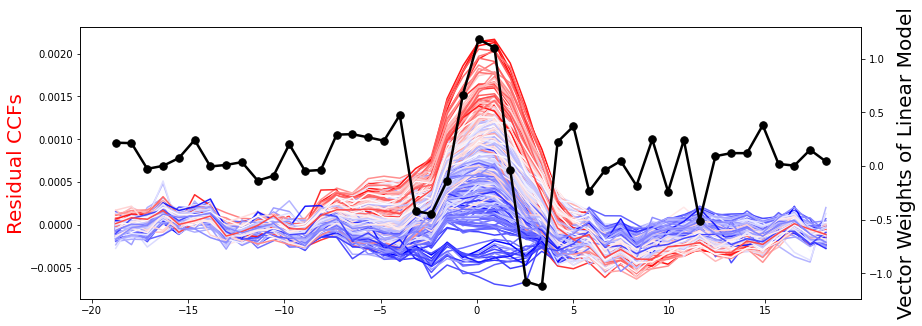}
    \caption{\bedit{Linear model input CCFs (before normalization) and corresponding vector weights 
    \textit{--} The Residual CCFs that were included in training set for our best linear model ($\alpha = 3.609$) are plotted in red, white, and blue where the color corresponds to the RV signal in the same way as Figure 4. The model input CCFs are normalized but we plot them before normalization here for visualisation purposes. In black, the learned vector weights are plotted. The weights show correlations with their neighbors (reflecting the $\approx$ 3 \kms\ HARPS-N instrumental profile) and show most information comes from points within 5 \kms of the line center.}}
    \label{fig:vector_weights}
\end{figure*}

\subsubsection{Linear model implementation}\label{lin_mod_impl}

We implemented our linear model in \texttt{scikit-learn}, an open-source library for ML in python \citep{scikit-learn}. Specifically, we performed a ridge regression which is equivalent to a linear regression with L2 regularization.

We performed random searches across the parameter space for $\alpha$ which determines the regularization strength. The values of $\alpha$ spanned 0 to \bedit{800 and significantly affect the model performance as illustrated in \textit{Figure \ref{fig:alphalinear}}. In Table \ref{harpsnvalcrossvaltablelinear}, the best model's performance across the validation and cross-validation sets is listed}. Each model was trained on a single CPU and the training time took $\sim$1 minute for the ten runs over which we average the predictions to compute our final stellar activity predictions.

\bedit{For our linear model, we extracted the vector weights and plot these alongside the input CCFs in \textit{Figure \ref{fig:vector_weights}}. In the bottom panel of \textit{Figure \ref{fig:vector_weights}}, neighboring weights appear correlated and the largest weights are concentrated in the center, which is expected from visually examining the input CCFs in the top panel of \textit{Figure \ref{fig:vector_weights}}.}

\begin{deluxetable}{l|b|b}[ht!]
\centering
\tablecaption{HARPS-N Validation and Cross-Validation\label{harpsnvalcrossvaltablelinear}}
\tablewidth{0.9pt}
\tablehead{\multicolumn{3}{c}{Linear model Best Hyperparameters}}
\startdata
\multicolumn{2}{l|}{Linear hyperparameters}       & \multicolumn{1}{c}{Best Model}  \\
\hline
\multicolumn{2}{l|}{$\alpha$}   & 3.609   \\  
\hline\hline
\multicolumn{3}{c}{ Validation Set Results} \\ [0.5ex]
\hline
\multicolumn{1}{l|}{Scatter} & \multicolumn{1}{c|}{Raw}  & \multicolumn{1}{c}{Corrected Data Using} \\ [0.5ex]
\multicolumn{1}{l|}{Metric} & \multicolumn{1}{c|}{Data} & \multicolumn{1}{c}{Best Model} \\
\hline
$\sigma_{SD}$ (\ms)  & 1.923 & 1.346 \\
$\sigma_{k \cdot MAD}$ (\ms) & 1.751 & 1.250 \\
$\sigma_{Percentile}$ (\ms)  & 2.083 & 1.430 \\
\hline\hline
\multicolumn{3}{c}{ Cross-Validation Set Results} \\ [0.5ex]
\hline
\multicolumn{1}{l|}{Scatter} & \multicolumn{1}{c|}{Raw}  & \multicolumn{1}{c}{Corrected Data Using} \\ [0.5ex]
\multicolumn{1}{l|}{Metric} & \multicolumn{1}{c|}{Data} & \multicolumn{1}{c}{Best Model} \\
\hline
$\sigma_{SD}$ (\ms)  & 1.828 &  1.085 \\
$\sigma_{k \cdot MAD}$ (\ms) & 1.744 & 1.085  \\
$\sigma_{Percentile}$ (\ms)  & 1.745 & 1.097
\enddata
\tablecomments{To find the best linear model architecture for the HARPS-N Observations, we performed random searches across the parameter space. The best linear model configuration and its corresponding reductions in RV scatter are listed here. This model was chosen as the final model based on its better performance across the validation and cross-validation sets. This model was thus used on the test set.}
\end{deluxetable}

\subsubsection{FC NN and CNN model implementation}

We implemented our FC NN and CNN models in TensorFlow, an open-source software library for ML algorithms \citep{abadi2016tensorflow}.

We used Stochastic Gradient Descent (SGD) with momentum to minimize the loss function over the training set. We performed random searches across the parameter space for the learning rate, weight decay, kernel size, filters, number of layers, and the number of epochs over the (cross-)validation set as listed in \textit{Table \ref{tab:cnn_fcnn_space}}. Each model was trained on a single CPU and the training time ranged from 5 minutes to 20 minutes depending on the complexity of the model. Across the 10 runs whose results we average, our best models took $\sim$ 5 and 8 minutes to train for the fully connected and convolutional architectures respectively. In \textit{Table \ref{harpsnvalcrossvaltablefc}} and \textit{Table \ref{harpsnvalcrossvaltable}}, the three best performing model hyperparameters are listed for the FC NN and CNN architectures respectively. Further, the final model architectures used for both the simulated SOAP 2.0 and HARPS-N Solar Data are listed in \textit{Figures \ref{fig:sllarchitectures}, \ref{fig:cnnarchitecture}}.

\begin{deluxetable*}{l|c|bb}[ht!]
\centering
\tablecaption{Random Search Hyperparameter Space\label{tab:cnn_fcnn_space}}
%\tablewidth{5000pt}
\tablewidth{0pt}
\tablehead{
\multicolumn{1}{c|}{Linear} & \colhead{Hyperparameter} & \multicolumn{2}{|c}{\hspace{1cm}Random Search \hspace{0.5cm} \textcolor{white}{.}} \\
\multicolumn{1}{c|}{hyperparameters} & \colhead{distribution}  & \multicolumn{1}{|c}{Space}}
\startdata
$\alpha$    & \bedit{Logarithmic}  & 0 - 1000 \\
\hline\hline
\multicolumn{1}{c|}{FC NN} & Hyperparameter & \multicolumn{2}{|c}{Random Search} \\ [0.5ex]
\multicolumn{1}{c|}{hyperparameters} & distribution  & \multicolumn{1}{|c}{Space} \\
\hline
Learning   rate    & $10^{-x}$ ($x$ is Uniform) & $10^{-4}-10^5$ \\
No.\ dense units  & Discrete & 50, 100, 200, 500, 1000, 2000  \\
No.\ dense layers & Discrete & 1, 2, 4, 8, 12, 16, 32    \\
Weight decay      & Logarithmic & $0.00001-0.1$  \\ 
\bedit{Epochs} & \bedit{Discrete} & 25, 30, 35, 40, 45, 50, 55, 60 \\
\hline\hline
\multicolumn{1}{c|}{CNN} & \colhead{Hyperparameter} & \multicolumn{2}{|c}{Random Search} \\
\multicolumn{1}{c|}{hyperparameters} & \colhead{distribution}  & \multicolumn{1}{|c}{Space}\\
\hline
Learning   rate    & $10^{-x}$ ($x$ is Uniform)  & $10^{-4}-10^5$ \\
Conv kernel size & Discrete      & 3, 5, 7 \\
No.\ conv filters & Discrete   & 8, 16, 32         \\
No.\ conv layers  & Discrete       & 2,4,6   \\
No.\ dense units  & Discrete  & 100, 200, 500, 1000       \\
No.\ dense layers & Discrete & 1, 2, 4, 6, 8     \\
Weight decay      & Logarithmic     & $0.0005-0.05$ \\
\bedit{Epochs} & \bedit{Discrete} & 50, 55, 65, 70, 80, 90, 100
\enddata
\tablecomments{To find the best hyperparameters across model architectures for both the SOAP2.0 and HARPS-N observations, we performed random searches across the parameter space. The ranges of the parameter space explored are the same for both datasets.  Convolutional layer parameters are denoted conv $\langle$parameter$\rangle$. Fully connected layer parameters are denoted dense $\langle$parameter$\rangle$. For FC NN and CNN models, we kept momentum = 0.9 which is a generally accepted value.}
\end{deluxetable*}

\begin{deluxetable}{l|b|bbb}[ht!]
\centering
\tablecaption{HARPS-N Validation and Cross-Validation\label{harpsnvalcrossvaltablefc}}
\tablewidth{1pt}
\tablehead{\multicolumn{5}{c}{FC NN Best Hyperparameters}}
\startdata
\multicolumn{2}{l|}{FC NN hyperparameters}   & Model   1 & Model   2 & Model   3  \\
\hline
\multicolumn{2}{l|}{Learning rate}   & 0.00161  & 0.00244  & 0.00135 \\  
\multicolumn{2}{l|}{No.\ dense units}  & 200 & 200    & 1000    \\
\multicolumn{2}{l|}{No.\ dense layers}   & 4  & 8  & 4   \\
\multicolumn{2}{l|}{Weight decay}    & 0.000100 & 0.000577 & 0.00010   \\
\hline\hline
\multicolumn{5}{c}{ Validation Set Results} \\ [0.5ex]
\hline
\multicolumn{1}{l|}{Scatter} & Raw  & \multicolumn{3}{c}{Corrected Data Using} \\ [0.5ex]
\multicolumn{1}{l|}{Metric} & Data & Model 1 & Model   2 & Model   3 \\
\hline
$\sigma_{SD}$ (\ms)  & 1.923 & 1.382  & 1.377    & 1.417    \\
$\sigma_{k \cdot MAD}$ (\ms) & 1.751 & 1.353  & 1.336    & 1.333    \\
$\sigma_{Percentile}$ (\ms)  & 2.083 & 1.425  & 1.335    & 1.447    \\
\hline\hline
\multicolumn{5}{c}{ Cross-Validation Set Results} \\ [0.5ex]
\hline
\multicolumn{1}{l|}{Scatter} & Raw  & \multicolumn{3}{c}{Corrected Data Using} \\ [0.5ex]
\multicolumn{1}{l|}{Metric} & Data & Model 1 & Model   2 & Model   3 \\
\hline
$\sigma_{SD}$ (\ms)  & 1.828 & 1.085  & 1.089    & 1.101    \\
$\sigma_{k \cdot MAD}$ (\ms) & 1.744 & 1.039  & 1.036    & 1.044    \\
$\sigma_{Percentile}$ (\ms)  & 1.745 & 1.064  & 1.038    & 1.101   
\enddata
\tablecomments{To find the best FC NN model architecture for the HARPS-N Observations, we performed random searches across the parameter space. The three best FC NN model configurations and their corresponding reductions in RV scatter are listed here. Although Model 1 and \bedit{2} perform similarly, Model \bedit{2} was chosen as the final model based on its marginally better performance across the validation set. Model \bedit{2} was thus used on the test set.}
\end{deluxetable}

\begin{deluxetable}{l|b|bbb}[ht!]
\centering
\tablecaption{HARPS-N Validation and Cross-Validation\label{harpsnvalcrossvaltable}}
\tablewidth{1pt}
\tablehead{\multicolumn{5}{c}{CNN Best Hyperparameters}}
\startdata
\multicolumn{2}{l|}{CNN hyperparameters}       & Model   1 & Model   2 & Model   3  \\
\hline
\multicolumn{2}{l|}{Learning rate}   & 0.016463   & 0.011615    & 0.0038415     \\
\multicolumn{2}{l|}{Conv kernel size}  & 11 & 11  & 9     \\
\multicolumn{2}{l|}{No.\ conv filters}  & 8  & 8   & 64    \\
\multicolumn{2}{l|}{No.\ conv layers}   & 1  & 1   & 1     \\
\multicolumn{2}{l|}{No.\ dense units}   & 100  & 2000  & 500   \\
\multicolumn{2}{l|}{No.\ dense layers}  & 1  & 1   & 4     \\
\multicolumn{2}{l|}{Weight decay}    & 0.033076   & 0.004515    & 0.000012362   \\
\multicolumn{2}{l|}{No.\ epochs}    & 75   & 100    & 80   \\
\hline\hline
\multicolumn{5}{c}{ Validation Set Results} \\ [0.5ex]
\hline
\multicolumn{1}{l|}{Scatter} & Raw  & \multicolumn{3}{c}{Corrected Data Using} \\ [0.5ex]
\multicolumn{1}{l|}{Metric} & Data & Model 1 & Model   2 & Model   3 \\
\hline
$\sigma_{SD}$ (\ms)  & 1.923 & 1.431      & 1.392  & 1.399  \\
$\sigma_{k \cdot MAD}$ (\ms) & 1.751 & 1.367      & 1.548  & 1.383  \\
$\sigma_{Percentile}$ (\ms)  & 2.083 & 1.376      & 1.354  & 1.267  \\
\hline\hline
\multicolumn{5}{c}{ Cross-Validation Set Results} \\ [0.5ex]
\hline
\multicolumn{1}{l|}{Scatter} & Raw  & \multicolumn{3}{c}{Corrected Data Using} \\ [0.5ex]
\multicolumn{1}{l|}{Metric} & Data & Model 1 & Model   2 & Model   3 \\
\hline
$\sigma_{SD}$ (\ms)  & 1.828 & 1.118      & 1.089  & 1.083  \\
$\sigma_{k \cdot MAD}$ (\ms) & 1.744 & 1.047      & 1.129  & 1.068  \\
$\sigma_{Percentile}$ (\ms)  & 1.745 & 1.016      & 1.123  & 1.027  
\enddata
\tablecomments{To find the best CNN model architecture for the HARPS-N Observations, we performed random searches across the parameter space. The three best CNN model configurations and their corresponding reductions in RV scatter are listed here. Model 3 was chosen as the best final model and used on the test set. Convolutional layer parameters are denoted conv $\langle$parameter$\rangle$.}
\end{deluxetable}

\subsection{Model Ensembling by Averaging}

After optimizing our hyperparameters for a specific architecture, we train 10 independent copies with different random parameter initializations. We then average the 10 outputs for all predictions to compute our results in \textit{Section \ref{results}}. This method of model averaging often improves performance. Across the input space, different versions of the same configuration may perform better or worse and this process averages out this difference in performance. This is especially important when the training set is small and we are at higher risk for overfitting. In addition, model averaging reduces the variance arising from randomness in parameter initialization and data ordering during training, making it easier to compare different architectures.

\begin{figure*}
    \epsscale{1.23}
    \plotone{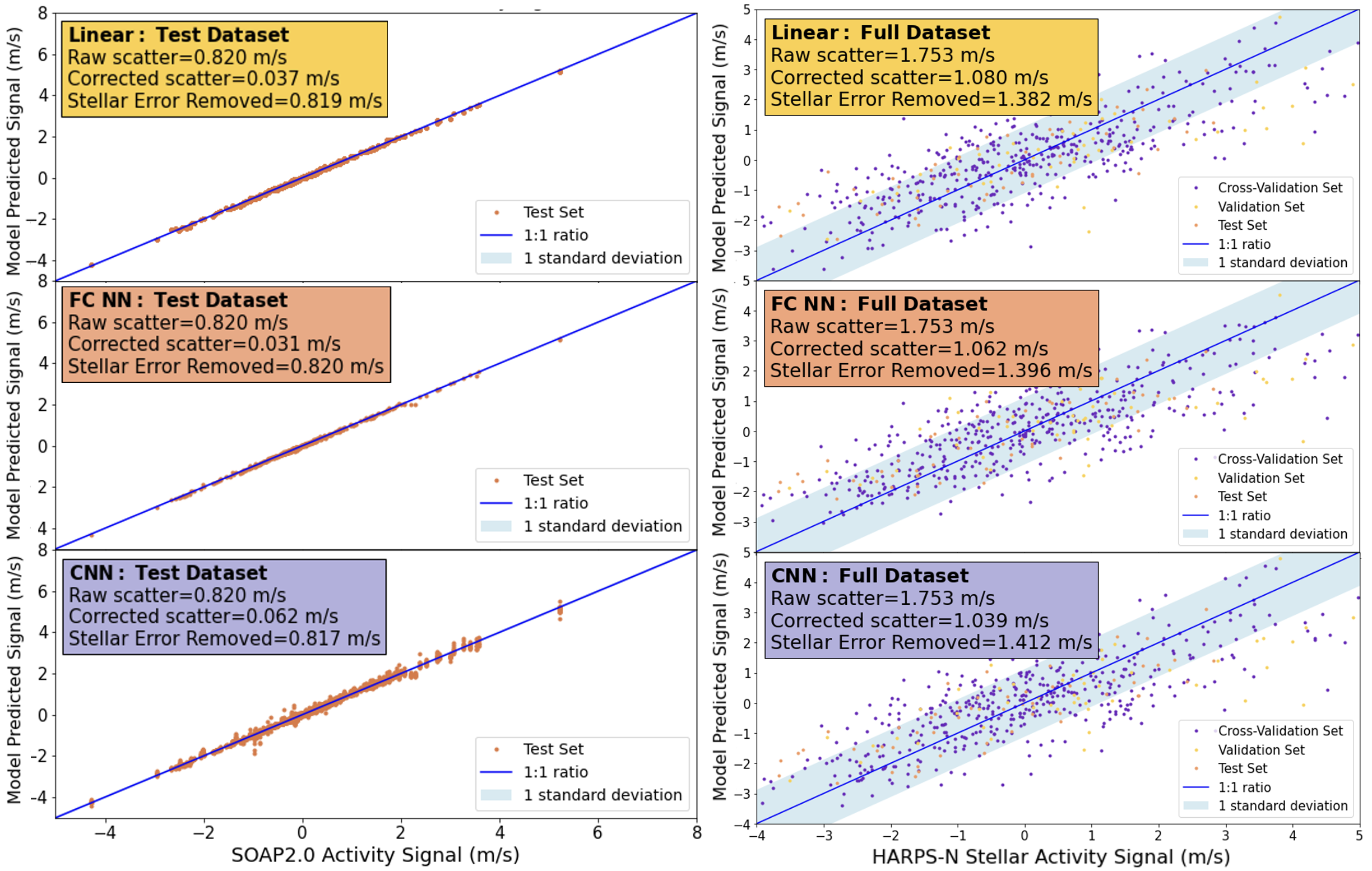}
    \caption{\color{new_color}{Linear, Fully Connected, and Convolution Neural Network Results for SOAP2.0 (column I) and HARPS-N Data (Column II). The scatter metric is standard $\sigma_{SD}$ for the SOAP 2.0 simulated data and $\sigma_{percentiles}$ for the non-Gaussian HARPS-N observations. The stellar error removed is the difference between the raw scatter and corrected scatter in quadrature. For the SOAP 2.0 simulated data, the FC NN model performs best across the test set reducing the raw scatter from 82.0 \cms\ to 3.1 \cms. For the HARPS-N Solar data, the CNN model marginally outperforms the FC NN architecture by reducing the RV scatter from \bedit{175.3} \cms\ to \bedit{103.9} \cms compared to 106.2 \cms\ for the FC NN across the full dataset. }}
    \label{fig:results_for_all6}
\end{figure*}

\section{Results}\label{results}

Here we report the results of our ML activity predictions. First we discuss the metrics we used to evaluate the performance and then we summarize how the different models performed on each dataset. 

\subsection{Performance metrics: $\sigma_{SD}$, $\sigma_{k\cdot MAD}$, and $\sigma_{Percentile}$}

After we finish optimizing our model parameters and hyperparameters on the training data, we evaluate our models' performance by characterizing the scatter of the ``corrected'' RVs, which we define as: 

\begin{equation}
    RV_{corrected} = RV_{raw} - RV_{predicted}
\end{equation}

\noindent where $RV_{raw}$ are the input RVs\footnote{These are the RVs in the heliocentric frame as calculated in Step 2 of \textit{Section \ref{harps_input}.}} without any activity corrections, and $RV_{predicted}$ are the predictions from our ML models. We introduce three metrics to characterize the scatter in the corrected RVs: $\sigma_{SD}$, $\sigma_{k\cdot MAD}$, and $\sigma_{Percentile}$. 

The first metric we calculate is the standard deviation of the corrected RVs, $\sigma_{SD}$. The standard deviation is given by:

\begin{equation}
        \sigma_{SD} = \sqrt{\frac{1}{M-1}\sum_{i=1}^{M}( RV_{corrected,i} - mean(RV_{corrected}))^{2}}
\end{equation}

\noindent where $M$ is the number of corrected RV observations. For well-behaved datasets like the SOAP 2.0 simulated data, using just the $\sigma_{SD}$ metric is sufficient to characterize the scatter. On the other hand, the HARPS-N dataset is more complex, so we introduce two new metrics in addition to $\sigma_{SD}$: $\sigma_{k\cdot MAD}$, based on the Median Absolute Deviation (MAD) and $\sigma_{Percentile}$. We introduce these additional two metrics for HARPS-N data because our data does not follow a normal distribution perfectly, and the presence of a few outlier datapoints in the HARPS-N dataset made it difficult to assess the model performance using only the $\sigma_{SD}$  metric. These new metrics are less sensitive to outliers than $\sigma_{SD}$ . The MAD metric is defined for a set of corrected RVs as: %predictions $\hat{Y} = \{\hat{y}_1, \hat{y}_2, ..., \hat{y}_n\}$ as:

    \begin{equation}
        {\rm MAD} = {\rm Median}(\mid RV_{corrected,i} - {\rm Median}(RV_{corrected})\mid)
    \end{equation}

%As mentioned above, ${\rm RMSE}_{\rm test}$ combines the predictions and the labels to provide one measure of performance. For the other two metrics, we take a slightly different approach where we compare the scatter from the raw HARPS-N activity signals to the scatter of the corrected signals (= labels - predictions). Rather than merely calculating the scatter by taking the standard deviation, 
%In addition to calculating the RMSE to assess model performance, we introduce two other metrics: the Median Absolute Deviation (MAD) and the half-68\%-interval. We use these other metrics because our data does not follow a normal distribution perfectly and these metrics are less sensitive to outliers than RMSE. 

%    \begin{equation}
%        {\rm MAD} (\hat{Y}) = {\rm Median}(\mid \hat{y}_i - {\rm Median}(\hat{Y})\mid)
%    \end{equation}

\noindent In other words, MAD takes the median of the data's absolute deviations around the data's median. To ease comparison with our other metrics like the standard deviation, we scale MAD by a factor $k$ such that:
\begin{equation}
    {\sigma_{k\cdot MAD}} = k \cdot {\rm MAD} 
\end{equation}
\noindent where $k$ depends on the type of distribution. For a normal distribution, $ k \approx 1.4826$ and we approximate our distribution as normal.
    
We call our final scatter metric $\sigma_{Percentile}$. We calculate this metric by computing: 
\begin{equation}
   \sigma_{Percentile} = \sfrac{1}{2} \left(RV_{\rm84th \%} - RV_{\rm 16th \%}\right)
\end{equation}

\noindent where $RV_{\rm84th \%}$ is the 84th percentile of the corrected RVs, and $RV_{\rm16th \%}$ is the 16th percentile of the corrected RVs. For a normal distribution, this is equivalent to calculating the standard deviation. However, our distribution of stellar activity signals is not perfectly normal and skewed by some of the outliers. Thus computing $\sigma_{Percentile}$ serves as a proxy for the standard deviation that is less sensitive to outliers.

\subsection{SOAP 2.0 Results}
For the simulated data using SOAP 2.0, our best performing models were the linear model and FC architecture which reduce the RV scatter, $\sigma_{SD}$, from 82.0 \cms\ to 3.7 \cms\ and  3.1 \cms\ across the test set respectively. These results are summarized in \textit{Figure \ref{fig:results_for_all6}} and \textit{Table \ref{soaptable}}. Thus, for the idealized case of simulated data, we can predict the stellar activity signal nearly exactly based on the shape changes in the normalized $\Delta$CCF.

\begin{deluxetable}{l|c|ccc}[ht!]
\centering
%\tablenum{1}
\tablecaption{SOAP 2.0 Results\label{soaptable}}
\tablewidth{1pt}
\tablehead{\multicolumn{5}{c}{ Test Set Results}}
\startdata
%\multicolumn{5}{c}{ Cross-Validation Set Results} \\
\multicolumn{1}{c|}{Scatter} & Raw  & \multicolumn{3}{c}{Corrected Data Using} \\  [0.5ex]
\multicolumn{1}{c|}{Metric} & Data & Linear Model & FC NN & CNN \\
\hline
$\sigma_{SD}$ (\ms)  &  0.820 & 0.037  & 0.031 & 0.062\\
\enddata
\tablecomments{We computed the standard deviation across the simulated stellar activity signals before applying any corrections (raw data) and then apply stellar activity corrections using all three model architectures. Their resulting reductions in scatter are listed across the test set. }
\end{deluxetable}

\begin{deluxetable}{l|b|bbb}[ht!]
\centering
%\tablenum{1}
\tablecaption{HARPS-N Results\label{harpsntable}}
\tablewidth{1pt}
\tablehead{\multicolumn{5}{c}{ Test Set Results}}
\startdata
%\multicolumn{5}{c}{ Cross-Validation Set Results} \\
\multicolumn{1}{c|}{Scatter} & Raw  & \multicolumn{3}{b}{Corrected Data Using} \\ [0.5ex]
\multicolumn{1}{c|}{Metric} & Data & Linear Model & FC NN & CNN \\
\hline
$\sigma_{SD}$ (\ms)  & 1.736 & 0.984   & 0.968  &  0.967  \\
$\sigma_{k \cdot MAD}$ (\ms) & 1.635 & 1.200   & 1.053  &  1.087  \\
$\sigma_{Percentile}$  (\ms) & 1.878 & 0.871   & 1.008  &  0.928  \\
\hline\hline
\multicolumn{5}{c}{Full Dataset Results} \\ [0.5ex]
\hline
\multicolumn{1}{c|}{Scatter} & Raw  & \multicolumn{3}{b}{Corrected Data Using} \\ [0.5ex]
\multicolumn{1}{c|}{Metric} & Data & Linear Model & FC NN & CNN \\
\hline
$\sigma_{SD}$ (\ms)  & 1.846 &  1.108  &  1.113 &  1.121 \\
$\sigma_{k \cdot MAD}$ (\ms) & 1.772 & 1.074   & 1.051  &  1.078  \\
$\sigma_{Percentile}$  (\ms) & 1.753 & 1.080   & 1.062  &  1.039  
\enddata
\tablecomments{We computed three different scatter metrics due to the slightly non-Gaussian nature of our data. We list these across the cross-validation results, the test set, and finally combine these corrected datasets (Full Dataset Results). }
\end{deluxetable}

\begin{figure*}
    \epsscale{1.1} %0.94 to fit with the next figure
    \plotone{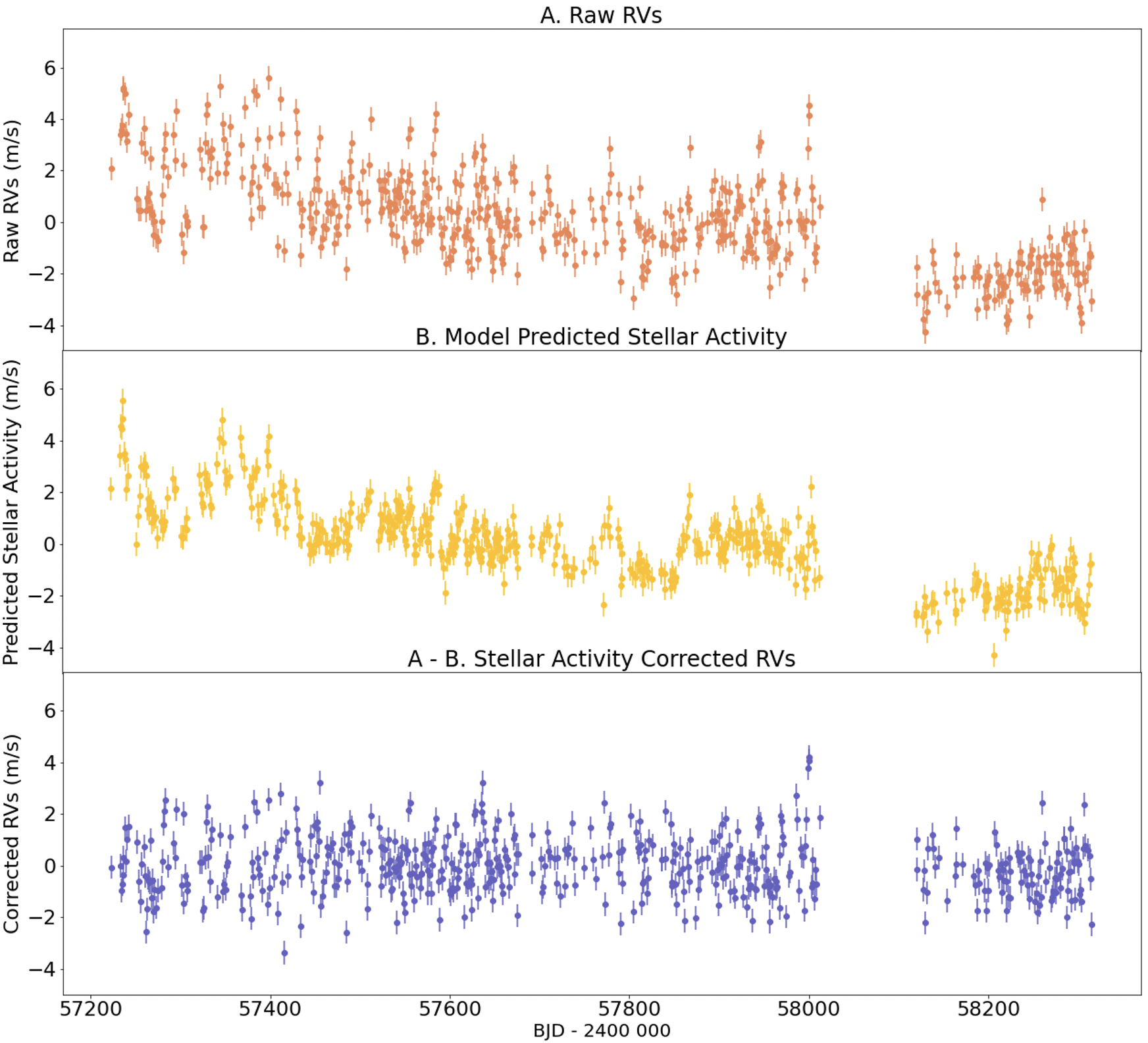}
    \caption{HARPS-N Solar Telescope Raw (A), Model Predicted (B) and Corrected (A - B) RVs over time. The stellar activity corrected RVs in the third panel are obtained by subtracting the predicted RVs (B) from the Raw RVs (A). The gaps in the observations $\sim$ 58100 days correspond to hardware downtime. }
    \label{fig:rvsvstime}
\end{figure*}

\begin{figure*}
    \epsscale{1.1} %0.94 to fit with the next figure
    \plotone{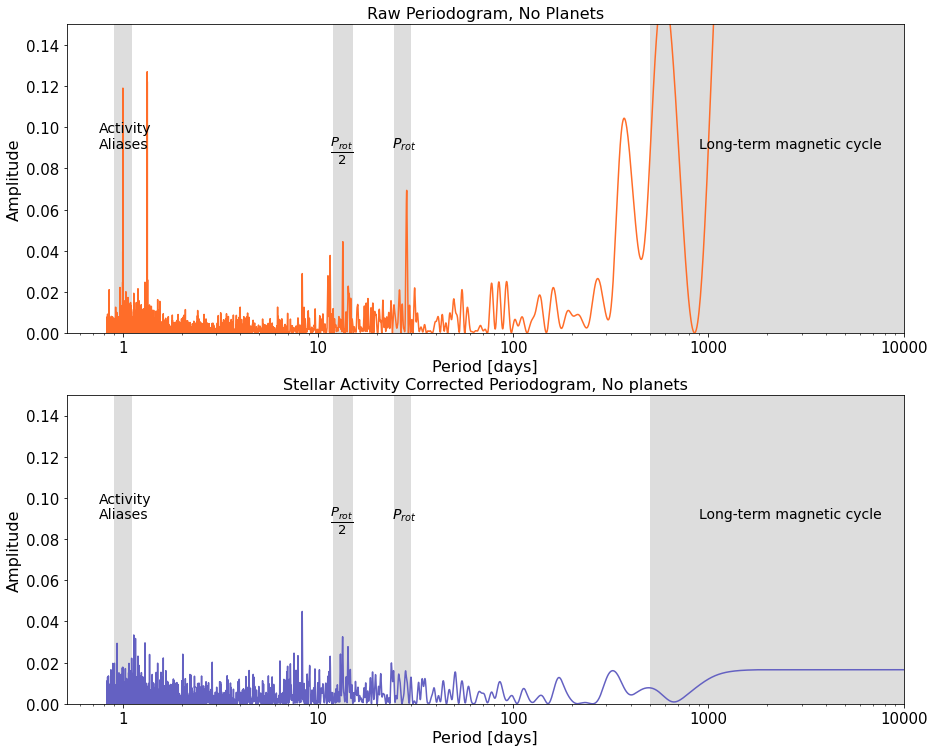}
    \caption{Periodogram: HARPS-N Solar Telescope Raw (a) and Corrected (b) RVs in Fourier space. The peaks in the top panel that correspond to stellar activity signals disappear in the bottom panel after applying the CNN model's stellar activity corrections. }
    \label{fig:correctedperiodogram}
\end{figure*}

\subsection{HARPS-N Results}

Our results across all three architectures are summarized in \textit{Figure \ref{fig:results_for_all6}} and \textit{Table \ref{harpsntable}}. Our best performing models were the FC NN and CNN (\textit{Figure \ref{fig:cnnarchitecture}}) which reduced the RV scatter,  $\sigma_{Percentile}$, from \bedit{175.3} \cms\ to \bedit{106.2} \cms and \bedit{103.9} \cms\ respectively across the full dataset. This remaining scatter is likely dominated by instrumental noise, not photon statistics.
Closely following the performance of the FC NN and CNN, our linear model reached a minimum scatter of \bedit{108} \cms\ across the full dataset. \bedit{From Table \ref{harpsntable}, we note that the overall reduction in scatter varies slightly across scatter metrics and methods.} Overall, these results suggest that \bedit{all three} model architectures match the structure of the HARPS-N Solar Data well, but the CNN model is potentially marginally more suitable.  The raw RVs, CNN predicted RVs, and CNN stellar activity corrected RVs are plotted over time in \textit{Figure \ref{fig:rvsvstime}}.

\subsubsection{Periodogram: Activity Signal Peaks Disappear}
We investigated the behavior of the raw and corrected HARPS-N RVs in the Fourier domain to see which signals are being removed to achieve this reduction in RMS scatter. \textit{Figure \ref{fig:correctedperiodogram}} shows the Lomb-Scargle Periodograms \citep{1976Ap&SS..39..447L, 1982ApJ...263..835S} of the RVs before and after applying the activity correction. To implement a generalized Lomb-Scargle Periodogram, we used the periodogram functions in \texttt{astropy.timeseries} \citep{2012cidu.conf...47V, 2015ApJ...812...18V}, where the periodograms are normalized according to the formalism in \citet{2009A&A...496..577Z}.

In the top panel of \textit{Figure \ref{fig:correctedperiodogram}}, the peaks at $\sim$ 25 days and $\sim$ 12 days correspond to the Sun's rotation period at the equator and half the rotation period respectively. The signal beyond $>$ 900 days is the long-term magnetic cycle. Lastly, the peaks at $\sim$ 1 day correspond to aliases from both the rotation period signals and the long-term magnetic cycle. After applying the corrections from our CNN, the periodogram of the corrected RVs no longer has peaks corresponding to these activity signals. Thus, the CNN is able to identify and remove the quasi-periodic variability at the stellar rotation period based only on shape changes in the $\Delta$CCF, and no information about when the observations were collected.

%\subsection{Using a Linear Model to Predict Radial Velocities}
%\subsection{Using more complex models to predict radial velocities}
%\subsubsection{Fully Connected Neural Network (FC NN)}
%\subsubsection{Convolutional Neural Network (CNN)}

\section{Discussion}\label{discussion}

\subsection{What is limiting our precision?}

Using machine learning, we were able to predict and remove stellar activity signals from HARPS-N Solar telescope observations and reduce the scatter in the measured RVs by about a factor of two from from \bedit{175.3} \cms\ to \bedit{103.9} \cms. While this improvement in RV precision is impressive and could increase our sensitivity to small planets if applied to observations of stars other than the Sun, our final scatter is still far greater than the roughly 10 \cms\ precision necessary to detect habitable-zone Earth analogs around Sun-like stars.  

What is limiting the precision of our activity-corrected HARPS-N RVs? One possibility is that our stellar activity corrections are not perfect, and the scatter is our corrected velocities is dominated by residuals stellar activity signals. However, we think that this is unlikely. We see no evidence for any quasi-periodic stellar activity signals in the periodogram of our corrected RVs (see \textit{Figure \ref{fig:correctedperiodogram}}), and our experiments with the SOAP 2.0 simulated data indicate that it is possible to achieve few \cms\ precision after modeling and removing stellar activity signals in a similar configuration. Certainly real data will have complications and subtleties that make stellar activity harder to correct than in our idealized SOAP 2.0 simulations, but it seems unlikely that these differences would cause our limiting precision to be 20 times greater. Several other analyses of the HARPS-N Solar Data seem to agree that, even when we successfully model activity at the rotation period using a variety of different techniques, there is still some other process limiting our RV precision \citep{2019ApJ...874..107M,2018A&A...620A..47D, 2020ApJ...888..117M}. 

It is likely that the remaining scatter in our corrected HARPS-N data is dominated by instrumental noise. While HARPS-N is highly stabilized, the instrument does experience slow drifts and requires frequent calibrations to ensure the accuracy of its wavelength solution. The quality of these wavelength solutions limits the precision of velocities measured by HARPS-N.  \citet{dumusque2020} report that wavelength solutions generated by the version of the DRS we use in this paper tend to change by about 74 \cms\ on day-to-day timescales, which could explain almost all of the scatter we see in our corrected HARPS-N solar velocities. If this is the case, then this technique could in principle yield more precise velocities when applied to data from newer stabilized spectrographs like ESPRESSO \citep{pepe2020}. 

%\textcolor{red}{neural network reduced the scatter in HARPS by 50\%. Can we do better? well, SOAP2.0 suggests that this phenomenone can describe this phenomenone completely. Hit the noise floor of HARPS, but what if we moved to ESPRESSO, EXPRES, NEID, MAroonX}

\subsection{Comparison to other methods}

Other common methods of reducing the RV scatter by characterizing and removing stellar activity signals include GPs. In a recent paper by \citet{langellier2020detection}, GPs \bedit{reduce the RMS scatter of the HARPS-N dataset to a similar reduction in RV scatter as our ML methods. One notable difference is that} we achieve this reduction in RMS scatter without using any information about the timing of the observations, potentially eliminating the need for high cadence sampling\footnote{Our method does of course require a rich dataset for training, but in principle the training observations could be taken with any cadence, as opposed to GPs, which often requires multiple observations per stellar rotation period to effectively model activity.}.

Another promising method for predicting stellar activity signals is to track the unsigned (unpolarised) magnetic flux as a proxy \citep{2020arXiv200513386H}. By estimating rotationally modulated RV variations and the unsigned magnetic flux daily over 8 years using spatially resolved SDO/HMI images, \citet{2020arXiv200513386H} showed that a simple fit with unsigned magnetic flux reduces rotationally modulated RV scatter by 62\% (a factor of 2.6 improvement). They successfully recovered planet semi-amplitudes of 0.3 \ms\ at orbital periods of $\sim$ 300 days. These numbers are not directly comparable to the work presented here because of different instrumental systematics and observational baselines; however, the improvement is of similar order. The authors note, however, that the unsigned magnetic flux is not yet measurable at high precision in slowly rotating, relatively inactive stars like the Sun. While this measurement is readily available for the Sun, making similar measurements for other stars will require pushing beyond the current state of the art in measuring Zeeman broadening from stellar spectra.

\bedit{Recently, \citet{CollierCameron2021} also explored the shape changes introduced in CCFs by computing an autocorrelation function that is invariant to translation (and thereby not sensitive to planetary reflex motion) but focused on stellar activity shape changes. In this analysis, the full 5 years of the HARPS-N dataset were used and injected planet signals of K = 0.4 \ms and periods ranging from 7 to 200 days were recovered. Since this analysis used the full 5 years and an older version of the DRS, these numbers are not directly comparable to our work. However, the improvement in RMS is similar, which further supports that these stellar activity-driven shape changes can serve as a useful indicator for the stellar RV contribution.}

\subsection{Implications for planet detection}\label{impl_planets}

To estimate the implications this ML method could have for planet detection, we injected a synthetic planet signal into both the full dataset of raw RVs \bedit{and raw CCFs} (\bedit{528} days of observations). \bedit{We then ran our full pipeline as described in detail in Sections \ref{inputrepresentation},  \ref{train_val_test}, and \ref{training_proc}}. We attempted to detect the signals in a Lomb-Scargle periodogram and assess their significance using Markov Chain Monte Carlo (MCMC).

\begin{figure*}
    \epsscale{1.15}
    \plotone{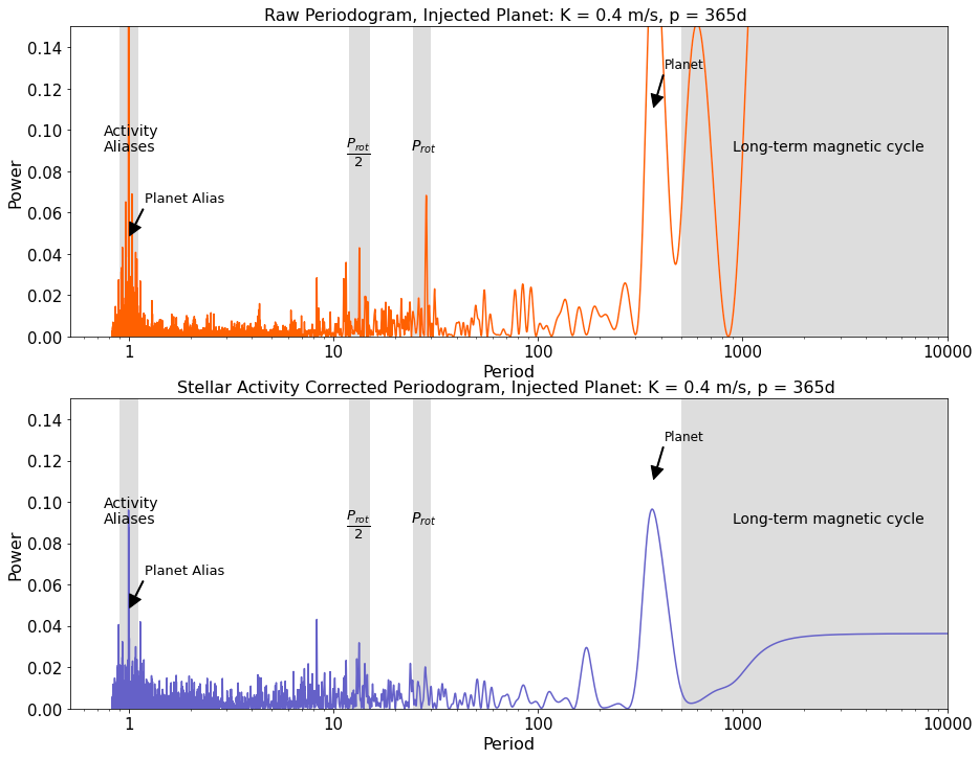}
    \caption{Periodogram, Planet Injection: K=0.3\ms, P = 365.24 d - In the top raw periodogram, the injected planet signal is distinct, but not the most prominent signal. Once these prominent stellar activities are corrected by CNN model, this planet (and its alias) becomes the most dominant signal (bottom panel). }
    \label{fig:injectedperiodogram}
\end{figure*}

\begin{figure*}
    \epsscale{1.1}
    \plotone{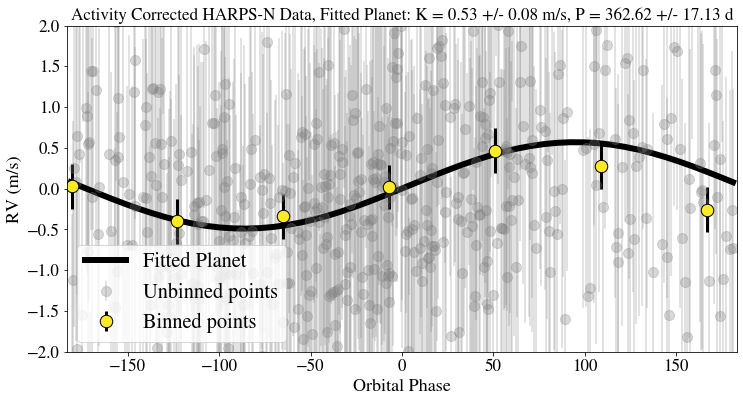}
    \caption{MCMC fitted Planet: K=$\textcolor{new_color}{0.53\pm0.07}$ \ms, P = $\textcolor{new_color}{362.69^{+9.26}_{-8.03}}$ days.
    The MCMC corner plot can be found in the \textit{Appendix, \textit{Figure \ref{fig:cornerplot}}}.}
    \label{fig:injectedmcmc}
\end{figure*}

%\subsubsection{Planet Injection: K=0.3 \ms, P = 365.24 d}

In \textit{Figure \ref{fig:injectedperiodogram}}, we show a periodogram of the HARPS-N RVs before and after activity corrections with a planet signal injected with a semi-amplitude of \bedit{0.4} \ms\ at a 1 year period (corresponding to a planet of \bedit{4.53}$M_{\Earth}$). In the periodogram of the uncorrected HARPS-N RVs (top panel), the injected signal is visible, but difficult to distinguish from other stronger peaks in the periodogram caused by stellar activity. However, in the periodogram of the corrected RVs (bottom panel), this planet signal is clearly the most prominent after we corrected for stellar activity signals. 

Using the periodograms to initialize our MCMC, we derived the phase-folded fit in \textit{Figure \ref{fig:injectedmcmc}} with K=$\textcolor{new_color}{0.53\pm0.07}$ \ms, P = $\textcolor{new_color}{362.69^{+9.26}_{-8.03}}$ days (\textit{Figure \ref{fig:cornerplot} in Appendix}). The MCMC uncertainties in this fit seem to be slightly underestimated, likely due to the non-Gaussian noise properties in the RVs.

The sensitivity of our method appears to compare favorably with the sensitivity of GP regression.  \citet{langellier2020detection} found that they would need 10-15 years of HARPS-N Solar data to detect an 0.5 \ms\ RV signal at a 225 day period with 5$\sigma$ confidence. We found that we can recover a 0.4 \ms\ injected signal with $\sim5\sigma$ confidence using only 3 years of HARPS-N Solar data.

\subsection{Future Work and Prospects for this Technique on Other Stars}\label{futurework}

To extend this method to other stars, we could take two different approaches. One approach would be to focus on one star at a time where we fit a simple (linear) ML model simultaneously with planet signals. This method has the advantage that it does not require the removal of astrophysical signals before fitting and that we would only need data for one star at a time. Some of the disadvantages would be that it could limit the complexity of the ML model and be more likely to overfit or have degeneracies where it fails to properly distinguish activity from planet signals, which is a common problem for other techniques like GPs. \bedit{Preliminary explorations of this type of simplified activity model has shown promising results on data from both HARPS-N (de Beurs et al. \textit{in prep}) and EXPRES \citep{zhao2022} on stars with between 25 and 100 observations.}

Another approach would be to train a more complex model on all stars observed by a given spectrograph simultaneously and predict stellar activity corrections for new stars (not included in the training set) based on the entire ensemble. The advantages would be that the larger training set would allow for more complex ML models that can predict the activity signals more accurately. However, this method has the disadvantage that the planet signals will need to be removed ahead of time. Some undetected planet signals will always remain in the data, meaning we would lack a perfect ``ground truth'' on which to train our models. Instead, we would have to hope that undetected signals would average out across the training set. In addition, the different rotation rates, spectral types, and inclination angles may be challenging to solve and require significantly more model complexity. Adding input features to our models, such as the $\textrm{log R}_{\textrm{HK}}^{'}$ or H$\alpha$ time series, stellar parameters like effective temperature and stellar radius, or stellar inclination angles derived from measurements of the projected rotational velocity and rotation period, may help our models make more accurate predictions.

In some ways, observations of stars at nighttime may be simpler to use as inputs to ML models. Unlike solar observations, nighttime observations have the following properties: 
\begin{enumerate}
    \item  Differential extinction is significantly less of a concern for observations of other stars at nighttime. Unlike the sun, the other stars that we observe are essentially point sources. Thus, differential extinction across the disc would not be resolved and induce significantly less systematic signals.
    \item There are some yearly effects on the CCFs of the sun that will not appear in stars observed at nighttime. In solar observations, the Full Width at Half Maximum (FWHM) of the CCF is modulated with 6-month and 1 year timescales. This phenomenon is due to the eccentricity of Earth's orbit, which causes the Earth's angular velocity about the Sun to vary annually and changes the relative angular velocity of the Sun's rotation that we observe. The changing relative rotational velocity affects our measurement of the rotational broadening of solar spectral features and therefore causes variations in the FWHM of the CCF \citep[See \textit{Figure 8a} in][] {2019MNRAS.487.1082C}. The six-month oscillation in the FWHM arises from the obliquity of the ecliptic plane relative to the solar equator. 
\end{enumerate}

On the other hand, nighttime observations will introduce new challenges of their own. For nighttime observations, we will not be able to average out granulation as well as for the Sun due to lower cadence observations. For other stars, The spectral lines also move across the detector due to barycentric velocity changes. In addition, observations of stars other than the Sun will often be photon-limited, unlike our solar observations in which photon noise is negligible. Noisier observations will make separating activity signals from true RV shifts more difficult. Nighttime RV observations are strongly heteroskedastic, unlike solar observations, due to factors like the observing conditions and different exposure times. Training a model may require more sophisticated weighting than we used in this work. In the future, this deep learning method could be applied to spectrographs like ESPRESSO (Pepe et al 2020 \textit{accepted}) and EXPRES \citep{2016SPIE.9908E..6TJ}, and might perform especially well on the extremely large data sets expected to be collected by HARPS-3 \citep{2016SPIE.9908E..6FT, 2018MNRAS.479.2968H}.

In future NN architectures, it may be advantageous to add pooling layers. Pooling layers take advantage of translational invariance in the input of CNNs. On the one hand, adding pooling layers could help for other stars where there may be slight shifts in the CFF due to undetected planets. On the other hand, pooling layers may prevent the detection of precise shifts in RV due to magnetic features. While our initial tests with pooling layers on solar data did not seem to help network performance, they may be helpful on more complex datasets. 

Other possible future applications of our ML stellar activity model include detecting young planets orbiting young stars. Young stars' high rotation rates and high levels of stellar activity make them especially complex, but simultaneously open the door to studying planetary formation and migration mechanisms. A machine-learning technique to remove stellar activity signals could open the door to measuring more masses and densities for transiting planets orbiting bright young stars \citep{mann2018,vanderburg2018,david2019,newton2019}. In this way, studying young planets around young stars is crucial to exoplanet demographic studies \citep{damasso2020gaps}.

\section{Conclusion}\label{conclusion}

Achieving the extreme RV precision necessary to detect long-period Earth-mass exoplanets requires mitigating stellar activity signals. These dominant stellar activity signals hide the $\sim$10 \cms\ signatures of Earth analog exoplanets and are difficult to remove due to their unpredictable time evolution and quasi-periodic nature. Current methods for mitigating stellar activity signals often rely on high cadence and carefully timed observations, but even with these methods, the detection of an Earth analog around a Sun-like will be very difficult \citep{langellier2020detection}. 

We have demonstrated a machine-learning approach to removing stellar activity signals that does not require the frequent sampling and timing information that other methods (like GP regression) depend on. By interpreting small shape changes in the stellar spectra induced by stellar activity, our ML models can predict and remove these dominant signals. So far, we have trained and tested our methods on simulated data and observations of the Sun, and plan to apply these or similar methods to observations of other stars in the future. For stars other than our Sun, we will not have as many spectra that can serve as a training set for a single star of interest but we may be able to train on an ensemble of other stars instead. We will make our code publicly available online to the exoplanet community\footnote{\url{https://github.com/zdebeurs/exoplanet-ml/tree/master/exoplanet-ml/rv_net}}. 

We developed and tested our methods on both simulated data (Monte Carlo generated using SOAP2.0; \citealt{2014ApJ...796..132D}) and solar observations from the HARPS-N Solar Telescope \citep{dumusque2015}. Our best performing model for simulated data was a FC NN which successfully reduced the scatter in simulated stellar activity signals from 82.0 \cms\ to 3.1 \cms\ across the test set. For the HARPS-N Solar observations, our best performing models, a FC NN and a CNN, both reduced the RV scatter from from \bedit{175.3} \cms\ to \bedit{103.9} \cms\ across 3 years of observations. When comparing our result to works that use GPs for HARPS-N observations \citep{langellier2020detection}, our FC NN and CNN models achieve similar (slightly better) precision than GP regression on the same data, \textit{without the need for high cadence sampling and timing information}.

We explored how much an activity correction like the one we have demonstrated on the Sun could improve the detectability of planets in similar datasets around other stars. We injected planet signals into our activity-corrected HARPS-N observations and were able to recover signals with semi-amplitudes down to 30 \cms, improving upon our detection limits in un-corrected observations by more than a factor of 2. However, these are not end-to-end injection/recovery tests and represent a best-case improvement to detection sensitivity. Future work will focus on investigating how to robustly prevent the algorithms from potentially confusing planetary and stellar activity signals. Nonetheless, these tests demonstrate that these advanced techniques could potentially pave the way to revealing previously hidden planets around our closest stellar neighbors.
\vspace{0.1in}

% restates abstract and goes into more detail
% main points
% sentence on each section
% NN can help with stellar activity
% this test -> that result
% experimented with injection recovery -> result

%===============================================================================

%\section*{Acknowledgements}
\vspace{0.1in}
\section*{Acknowledgements}
\bedit{We thank the anonymous referee for a constructive and detailed report, which helped us clarify key points and improve this manuscript.}

We thank Ellen Price for invaluable assistance with python environments. We acknowledge helpful conversations and feedback from George Dahl and members of Dave Latham's Coffee Club. The HARPS-N project has been funded by the Prodex Program of the Swiss Space Office (SSO), the Harvard University Origins of Life Initiative (HUOLI), the Scottish Universities Physics Alliance (SUPA), the University of Geneva, the Smithsonian Astrophysical Observatory (SAO), and the Italian National Astrophysical Institute (INAF), the University of St Andrews, Queen's University Belfast, and the University of Edinburgh. 

ZLD acknowledges the generous support from the UT Office of Undergraduate Research Fellowship, the TIDES Advanced Research Fellowship, Dean’s Scholars, and the Junior Fellows Honors Program. ZLD and AV acknowledge support from the TESS Guest Investigator Program under NASA grant 80NSSC19K0388. AV's work was partially performed under contract with the California Institute of Technology (Caltech)/Jet Propulsion Laboratory (JPL) funded by NASA through the Sagan Fellowship Program executed by the NASA Exoplanet Science Institute.
XD is grateful to the Branco-Weiss Fellowship for continuous support. This project has received funding from the European Research Council (ERC) under the European Union’s Horizon 2020 research and innovation programme (SCORE grant agreement No 851555).  ACC acknowledges support from the Science and Technology Facilities Council (STFC) consolidated grant number ST/R000824/1 and UKSA grant ST/R003203/1. This work was performed under contract with the California Institute of Technology (Caltech)/Jet Propulsion Laboratory (JPL) funded by NASA through the Sagan Fellowship Program executed by the NASA Exoplanet Science Institute (R.D.H.). R.D.H. is funded by the UK Science and Technology Facilities Council (STFC)'s Ernest Rutherford Fellowship (grant number ST/V004735/1). MPi acknowledges financial support from the ASI-INAF agreement n. 2018-16-HH.0. AM acknowledges support from the senior Kavli Institute Fellowships.

%SU, FP, FB, DS, CL, DE, M.Marmier, and M.Mayor acknowledge financial support from the Swiss National Science Foundation (SNSF) in the frame work of the National %Centre for Competence in Research PlanetS. DE acknowledges financial support from the European Research Council (ERC) under the European Union’s Horizon 2020 research and innovation program (project {\sc Four Aces}; grant agreement 724427).
%NN acknowledges partial supported by JSPS KAKENHI Grant Number JP18H01265 and JST PRESTO Grant Number JPMJPR1775.
%We made use of the Python programming language \citep{Rossum1995} 
%and the open-source Python packages
%\textsc{numpy} \citep{vanderWalt2011}, 
%{\scshape scipy} \citep{Jones2001}, 
%{\scshape matplotlib} \citep{Hunter2007}, 
%\textsc{emcee} \citep{Foreman-Mackey2013}, 
%{\scshape george} \citep{Ambikasaran2014}, 
%and
%\textsc{celerite} \citep{Foreman-Mackey2017}.
%{\scshape rebound} \citep{rebound},
%and {\scshape reboundx}. 

%{\scshape corner} \citep{Foreman-Mackey2016}, 
%{\scshape seaborn} (\url{https://seaborn.pydata.org/index.html}),
%===============================================================================

%===============================================================================
\facility{HARPS-N Solar Telescope, SDO}

\textit{Software:} numpy \citep{np}, matplotlib \citep{plt}. Tensorflow, SOAP2.0, Astropy \citep{astropy:2018}, scipy \citep{2020SciPy-NMeth}

\bibliography{bibliography}{}
\bibliographystyle{aasjournal}

\iffalse

\fi

\appendix
\vspace{0.1in}

%\section{Relationship between capacity and overfitting}

%\begin{figure*}[ht!]
%\centering
%	\epsscale{0.8}
%	\plotone{generalization_error.png}
%    \caption{Typical relationship between capacity and error
%    \textit{--} adapted from \citep{Goodfellow-et-al-2016}}
%    \label{fig:goodfellow}
%\end{figure*}   

\begin{figure*}[h!]
    \epsscale{1.25}
    \plotone{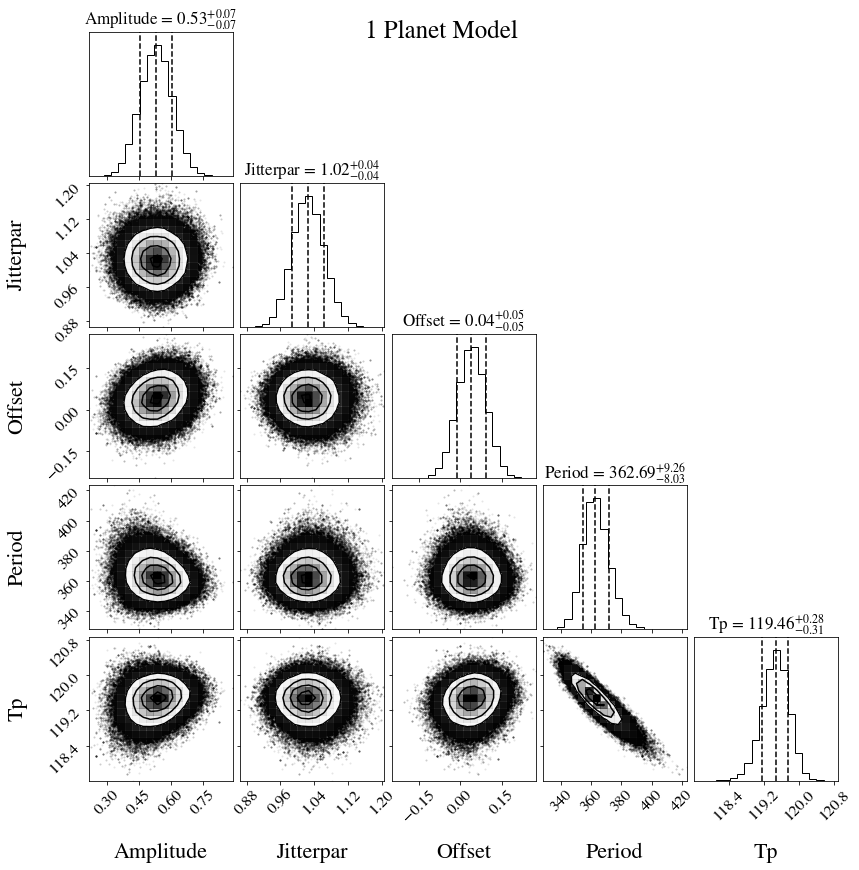}
    \caption{Corner Plot for MCMC fitted Planet:  K=$\textcolor{new_color}{0.53\pm0.07}$ \ms, P = $\textcolor{new_color}{362.69^{+9.26}_{-8.03}}$ days}
    \label{fig:cornerplot}
\end{figure*}

%\begin{figure*}
%    \epsscale{1.1}
%    \plotone{mcmcfit_p_1.3d_0.318ms_planets.png}
%    \caption{MCMC fit Planet II}
%    \label{fig:architecutresults}
%\end{figure*}

\end{document}